\documentclass[11pt,a4paper,onecolumn,nofootinbib]{revtex4}

\usepackage[cp1250]{inputenc}
\usepackage{graphicx} 
\usepackage{subfigure}
\usepackage{geometry}
\usepackage{amsmath}
\usepackage{multirow}

\usepackage{color} 

\newgeometry{tmargin=2cm, bmargin=2cm, lmargin=2cm, rmargin=2cm}

\begin{document}

\title{Complex Correlation Approach for High Frequency Financial Data}
\date{\today}
\author{Mateusz Wilinski\textsuperscript{1,2}}
\email{mateusz.wilinski@fuw.edu.pl}
\author{Yuichi Ikeda\textsuperscript{3}}
\author{Hideaki Aoyama\textsuperscript{4}}
\affiliation{\textsuperscript{1}Faculty of Physics, University of Warsaw, Pasteura 5, 02-093 Warsaw, Poland\\
\textsuperscript{2}Scuola Normale Superiore, piazza dei Cavalieri 7, 56126 Pisa, Italy\\
\textsuperscript{3}Graduate School of Advanced Integrated Studies in Human Survivability, Kyoto University, Kyoto, Japan\\
\textsuperscript{4}Graduate School of Science, Kyoto University, Kyoto, Japan}

\begin{abstract}
We propose a novel approach that allows to calculate Hilbert transform based complex correlation for unevenly spaced data.
This method is especially suitable for high frequency trading data, which are of a particular interest in finance.
Its most important feature is the ability to take into account lead-lag relations on different scales, without knowing them in advance.
We also present results obtained with this approach while working on Tokyo Stock Exchange intraday quotations.
We show that individual sectors and subsectors tend to form important market components which may follow each other with small but significant delays.
These components may be recognized by analysing eigenvectors of complex correlation matrix for Nikkei 225 stocks.
Interestingly, sectorial components are also found in eigenvectors corresponding to the bulk eigenvalues, traditionally treated as noise.
\end{abstract}

\maketitle

\section{INTRODUCTION}

Describing interactions between parts of a system is a crucial element of the complexity science.
World Wide Web, human brain, transportation networks, these are just few complex systems that attract the attention of researchers.
The main tools used to investigate connections' properties in highly coupled systems are complex networks \cite{boccaletti2006complex}.
For many complex systems described with networks, it is clear what kind of relation is represented by a particular edge.
In social networks, for example, it may be a friendship or a common work place.
In other cases, like scientific collaboration or city communication, we can even assign a weight to each connection \cite{barrat2004architecture}.
There are, however, complex systems for which the relationship between any two elements is implicit.
It is the case when we observe only the evolution of the system and at the same time we are unable to track the actual interactions and the way parts of the system influence each other.
This applies to, among others, brain activity and financial markets.
The latter will be the main topic of this paper.

In order to find the real structure of a system with hidden dependencies we need to analyse the observable outcome of system dynamics.
For financial markets this outcome might be stock quotations and for brain activity it may be electroencephalography (EEG) signal.
It is a common practice to use different measures that would be able to indicate which parts of the system are connected based on their signal similarity.
In finance, the most popular measure is the Pearson correlation coefficient which proved to be successful in uncovering hierarchical structure of the market \cite{mantegna1999hierarchical}.
In other areas, like in mentioned EEG signal, it is often needed to develop other methods that would suit the specifics of the data \cite{kaminski1991new}.

Using correlation as a measure of dependency in financial data has several drawbacks.
First, the correlation does not imply causation and its high values may be a consequence of indirect relations.
Some papers tried to deal with these problems by using more sophisticated methods, like partial correlation \cite{kenett2010dominating} or mutual information \cite{fiedor2014networks}.
Second, price movements of related stocks may be desynchronized.
In other words, one stock may lead the other one, which will be delayed.
Traditional Pearson correlation will not be able to find this kind of lead-lag relations.
This problem was approached by analysing shifted correlations \cite{huth2014high,kullmann2002time}.
Unfortunately, this methodology may be inefficient when dealing with high number of signals, since the delays in financial data may be different across different pairs of stocks.
Taking into account that in both \cite{huth2014high,kullmann2002time} around $40$ lags were analysed, it would mean estimating around $10^6$ correlations in case of our data.
Moreover, this number grows like $N^2$.
In this paper we propose a different approach, suited for high frequency financial data, which will take into account delays and do it in an efficient way.
Furthermore, we will use this approach to analyse intraday quotations data from Tokyo Stock Exchange (TSE).

\section{DATA}

The data used in this paper consist of all the trades and all the best bid and best ask offers submitted to the TSE in the period between 1 January and 31 December 2014.
We focused on the stocks that form the \textit{Nikkei 225}, the most important stock market index for the TSE and one of the most significant indexes in Asia.
We choose 222 stocks that were components of Nikkei in 2014 and were traded on every analysed day.
For each stock $i$ we calculated the mid price:
\begin{equation}
P_i(t) = \frac{A_i(t) + B_i(t)}{2},
\end{equation}
where $A_i(t)$ and $B_i(t)$ are respectively best ask and best bid for $i$th stock.
From mid prices we get the logarithmic prices:
\begin{equation}
p_i(t) = \log(P_i(t)),
\end{equation}
where $p_i(t)$ is a step function with steps at each quotation.
We also grouped all the stocks into sectors and subsectors, according to the official Nikkei 225 website\footnote{http://indexes.nikkei.co.jp/en/nkave}.
See Table \ref{tab:colors} for the names of sectors and subsectors, as well as the assigned colors and shapes used in further graphs.
The chosen data subset consist of more than $2 \cdot 10^9$ quotations, however, $90\%$ of them does not change the mid price.
The remaining part is around $1.3 \cdot 10^8$ significant quotations with the least actively traded stock having $1778$ data points and the second least $8632$ data point.
It should also be pointed out that the resolution of timestamps is one seconds, although, we know the precise order of quotes if they appear in the same second.

\section{METHODOLOGY}

A nonparametric methodology of calculating covariance and correlation matrix for multivariate data was proposed and described in details in \cite{malliavin2002fourier}.
It can be used with unevenly spaced time series and it already showed some meaningful results for financial data \cite{iori2007weighted}.
It was also shown to outperform other methods when dealing with finite samples \cite{nielsen2008finite}.
Thorough study showed as well, that it is almost unbiased in the presence of market microstructure noise \cite{mancino2008robustness}.
The idea behind this method is based on Fourier transformation.
Its major advantage is that it does not require any modifications in raw data.
We will describe the way this approach is used with empirical data.
More technical details can be found in \cite{malliavin2002fourier}.
Lets denote the analysed price process as $p_i(t) = \log P_i(t)$.
Additionally, let us assume that this process is well defined on the time interval $[0,T]$.
First, we rescale mentioned time interval into $[0, 2\pi]$.
Second, we need to calculate Fourier coefficients of $dp_i(t)$, which are defined as follows:
\begin{eqnarray*}
a_k(dp_i) = \frac{1}{\pi} \int_0^{2\pi} \cos (kt) dp_i(t), \\
b_k(dp_i) = \frac{1}{\pi} \int_0^{2\pi} \sin (kt) dp_i(t).
\end{eqnarray*}
\noindent Assuming that $p_i(t)$ is a step function and $t_m$ is the time of the $m$th quotation, the above integral becomes a summation:
\begin{eqnarray*}
\label{eq:fou_coef}
a_k(dp_i) = \frac{p_i(t_N) - p_i(t_1)}{\pi} - \frac{1}{\pi} \sum_{m=1}^{N-1} p_i(t_m) \left( \cos(kt_{m+1}) - \cos(kt_m) \right), \\
b_k(dp_i) = -\frac{1}{\pi} \sum_{m=1}^{N-1} p_i(t_m) \left( \sin(kt_{m+1}) - \sin(kt_m) \right),
\end{eqnarray*}
\noindent where $N$ is the number of quotations, $t_1 = 0$ and $t_N = 2\pi$.
From these coefficients, we can obtain the log-returns variance-covariance matrix $\Sigma_{ij}$ as a function of time.
In our case though, we only want the average covariance on a particular time period.
For this purpose, we only need the constant factor of the covariance Fourier representation, because other parts will reduce to zero in the process of averaging.
A precise derivation of this constant part may be found in \cite{malliavin2002fourier}.
We will only present the final result:
\begin{equation}
\label{eq:fou_cov}
a_0(\Sigma_{ij}) = \lim_{\tau \rightarrow 0} \frac{\pi\tau}{T} \sum_{k=1}^{T/2\tau} \left[ a_k(dp_i) a_k(dp_j) + b_k(dp_i) b_k(dp_j) \right],
\end{equation}
\noindent where $\tau$ determines the highest wave harmonic $\frac{T}{2\tau}$.
In the case of our TSE data the minimum value of $\tau$ could be one second, because this is the timestamp resolution.
However, after analysing the distribution of time intervals between quotations, we decided to set $\tau = 1$ minute.
This is consistent with findings from \cite{nielsen2008finite,mattiussi2011comparing} where it was shown that this is the most efficient scale for Fourier estimator.
Moreover, as shown in \cite{mancino2008robustness}, $\tau = 1$ minute gives the best balance between the problem of finite sample and asynchronicity.
It should also be pointed out that taking a specific $\tau$ is not in any way equivalent with aggregating data into evenly spaced time series and we still use all the data points.
Finally we can compute the covariance matrix estimator:
\begin{equation}
\hat{\sigma}_{ij}^2 = 2 \pi a_0(\Sigma_{ij}),
\label{eq:cov_cov}
\end{equation}
\noindent and further also the correlation matrix:
\begin{equation}
\rho_{ij} = \frac{\hat{\sigma}_{ij}^2}{\hat{\sigma}_{ii} \cdot \hat{\sigma}_{jj}}.
\label{eq:cov_cor}
\end{equation}
\noindent An important advantage of this estimator is that it gives accurate estimates for step function type of data.
This makes it particularly suited for financial data.

We can now use all the tick data available, but the problem of lead-lag relationships still holds.
To overcome it, we will adopt and adjust to our needs ideas from the \textit{Complex Hilbert Principal Component Analysis} (CHPCA) \cite{horel1984complex}, mainly the Hilbert Transform.
The CHPCA was used primarily in geophysics and meteorology.
Recently, it also proved to be effective with financial data \cite{arai2015dynamic,vodenska2016interdependencies}.
The previous research, however, was made using daily data and was applicable only to synchronous data with constant step.
Although it makes sense to suspect different countries indexes to have lead-lag relations that would span for periods longer than one day, it is rather not the case of stocks from the same market.
The increased activity of intraday traders and hedge funds made the market much more efficient and synchronized \cite{toth2006increasing}.
As a result, one needs to use high frequency data in order to find out which stocks are leading others.
Since we do not want to aggregate the data, which may lead to data loss, we would like to propose a way to use the Hilbert transform together with the Fourier algorithm described before.
We show that this may be done in a purely analytical way which not only allows us to use raw data, but also makes the procedure numerically efficient.

Hilbert transform is a linear operator that does not change the time domain of a transformed function.
A formal definition of this operator is as follows:
\begin{equation}
\label{eq:hil}
H(Z,t) = p.v. \frac{1}{\pi} \int_{-\infty}^\infty \frac{Z(s)}{t-s} ds,
\end{equation}
where $p.v.$ stands for Cauchy principal value and $Z$ is some process that we want to transform.
Hilbert transform, at particular time $t$, codes information about the future and past movement of transformed series.
For this reason it is useful to use it when looking for correlations among different signals, especially if we expect that the relation may be lagged, but we don't know the specific lag-value or we know that this lag may vary over time.
In order to keep the information about the time series value and its Hilbert transform, we produce a complexified time series.
This new time series has complex values with the real part being the original value and the imaginary part being the Hilbert transform.
\begin{equation}
\label{eq:comp_ser}
\hat{Z}(t) = Z(t) + i H(Z,t).
\end{equation}
Such time series may now be used in order to obtain complex correlations, which should contain both the information about immediate and lagged correlations.
Using Eq. (\ref{eq:hil}) to obtain Hilbert transform of a time series, may be very difficult.
Our goal, however, is to find the correlation matrix of complexified time series $\hat{Z}(t)$ and, as we will show, we do not need to know an explicit form of $H(Z,t)$.
In our case $Z(t) = dp_i(t)$ and we already know how to obtain its Fourier coefficients that lead to estimating its correlation matrix.
Moreover, Hilbert transform is additive and can be easily calculated for trigonometric functions, namely:
\begin{eqnarray*}
H(\sin(\cdot),x) = -\cos(x), \\
H(\cos(\cdot),x) = \sin(x).
\end{eqnarray*}
As a result, if we know Fourier coefficients of a given process we can easily calculate them also for its Hilbert transform:
\begin{eqnarray*}
a_k(H(Z)) = -b_k(Z), \\
b_k(H(Z)) = a_k(Z).
\end{eqnarray*}
This way, we can easily obtain Fourier coefficients of $\hat{Z}(t)$:
\begin{eqnarray*}
a_k(\hat{Z}) = a_k(Z) + i a_k(H(Z)) = a_k(Z) - i b_k(Z), \\
b_k(\hat{Z}) = b_k(Z) + i b_k(H(Z)) = b_k(Z) + i a_k(Z).
\end{eqnarray*}
Finally we can use it with Eq. (\ref{eq:fou_cov}) and obtain:
\begin{equation}
\label{eq:fou_hil_cov}
a_0(\Sigma_{ij}) = \lim_{\tau \rightarrow 0} \frac{\pi\tau}{T} \sum_{k=1}^{T/2\tau} \left[ (a_k(dp_i) - i b_k(dp_i)) (a_k(dp_j) + i b_k(dp_j)) + (b_k(dp_i) + i a_k(dp_i)) (b_k(dp_j) - i a_k(dp_j)) \right].
\end{equation}
Combining Eq. (\ref{eq:fou_coef}), (\ref{eq:fou_hil_cov}), (\ref{eq:cov_cov}), (\ref{eq:cov_cor}) allows us to calculate the complex correlation matrix directly, with no numerical approximations.
The precision of this estimate is limited only by the finite sample and data timestamp resolution.
In particular, including Hilbert transformation part is done in an entirely analytical way.
As a result, the precision of our estimation is the same as for the Fourier approach, namely it is proportional to $N^{-\frac{1}{2}}$.
It is also worth mentioning that the number of parameters needed to estimate complex correlation with our method is linear in $N$.
Basically we only need to obtain the parameters of the Fourier correlation estimator.
In contrast, previous works on complex correlation of daily financial data \cite{arai2015dynamic,vodenska2016interdependencies} required $O(N^2)$ estimates.

Each element $\rho_{kl}$ of the complex correlation matrix has the form:
\begin{equation}
\label{eq:comp_corr}
\rho_{kl} = s_{kl} e^{-i \theta_{kl}},
\end{equation}
where $s_{kl}$ is the magnitude of correlation and the $\theta_{kl}$ is the phase, which indicate the lead-lag relation between the stock $k$ and the stock $l$.
The last part of the data analysis done in this paper is the eigendecomposition of the correlation matrix.
This is often used as a tool of the random matrix theory for financial data \cite{laloux1999noise,plerou1999universal,utsugi2004random}.
Since the complex correlation matrix $\rho$ is Hermitian, we can write it in a following form:
\begin{equation}
\label{eq:decom}
\rho = \sum_{i=1}^{N} \lambda_i V^{(i)} V^{(i) \dagger},
\end{equation}
where $\lambda_i$ is the $i$th eigenvalue, $V^{(i)}$ is the corresponding eigenvector and $\dagger$ denotes a complex conjugate.
The complex principal components $CP_i$ may be derived from eigenvectors according to the equation:
\begin{equation}
\label{eq:comp}
CP_i(t) = \sum_{j=1}^{N} dp_j(t) V_j^{(i)},
\end{equation}
where $dp_j$ is in our case the process of log returns.
The important thing is that $V_j^{(i)}$ is actually the complex correlation between $j$th time series and the $i$th complex principal component.
The complex form of this correlation informs us not only about which stocks, or groups of stocks, are closely related with particular component but also about the lead lag relation between them.
We will exploit it in the empirical analysis of financial data.

\section{RESULTS}

\begin{figure}
\centering
\includegraphics[width=0.9\textwidth]{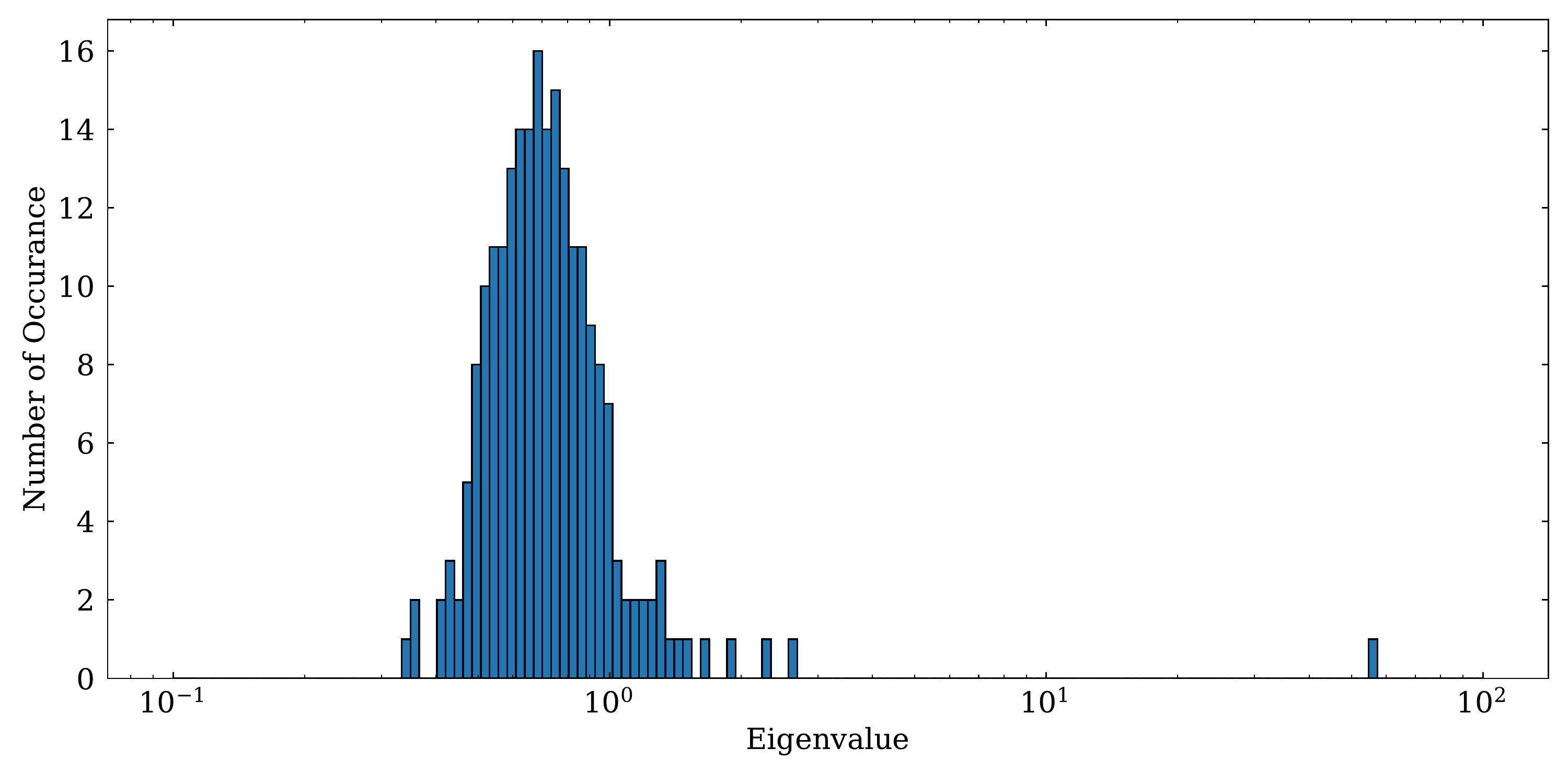}
\caption{Histogram of the frequency of eigenvalues of complex correlation matrix for TSE intraday data (2014).}
\label{fig:eigen_hist}
\end{figure}

\begin{figure}
    \centering
	\subfigure[ Eigenvector corresponding to the largest eigenvalue.]{\includegraphics[width=0.49\textwidth]{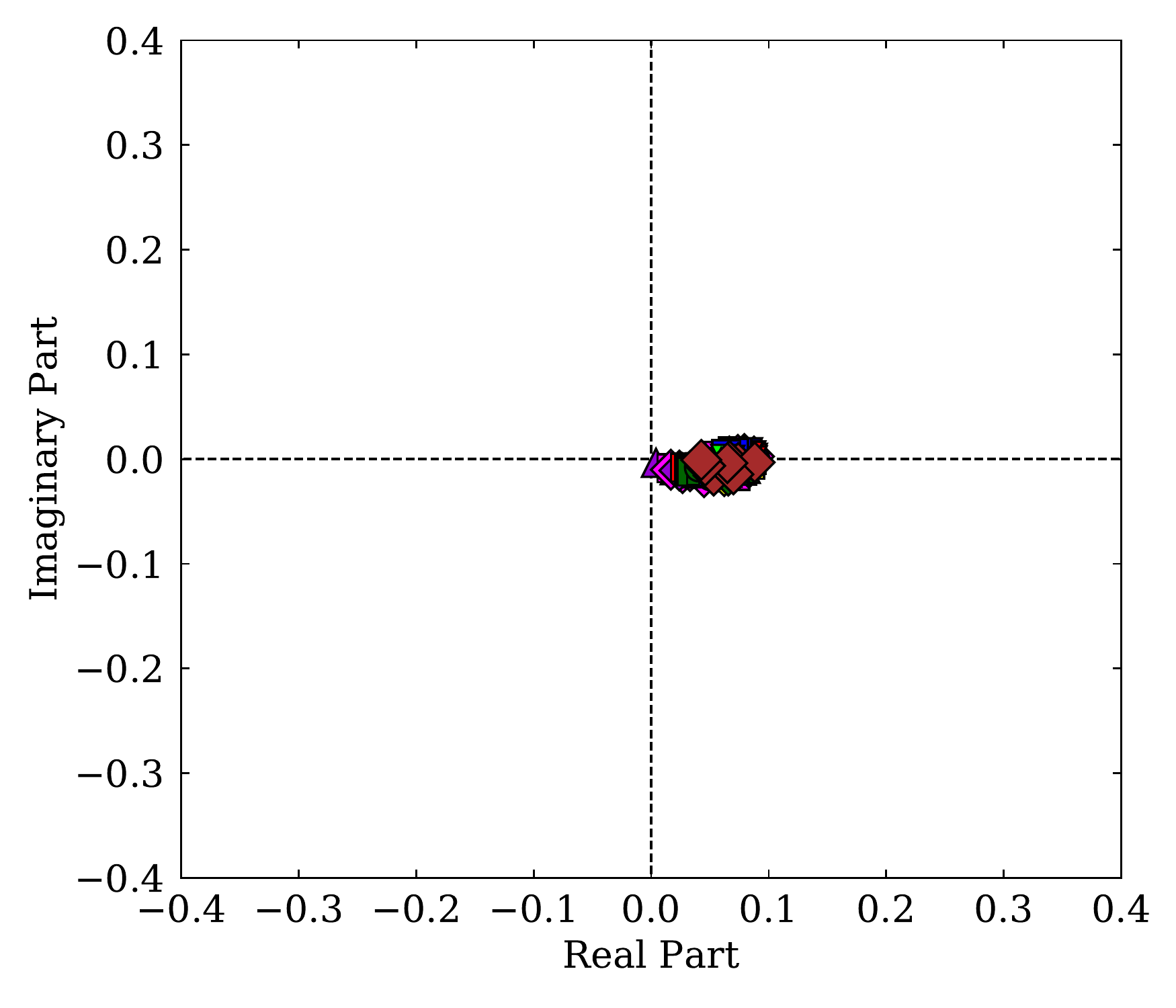}}
	\subfigure[ Eigenvector corresponding to the second largest eigenvalue.]{\includegraphics[width=0.49\textwidth]{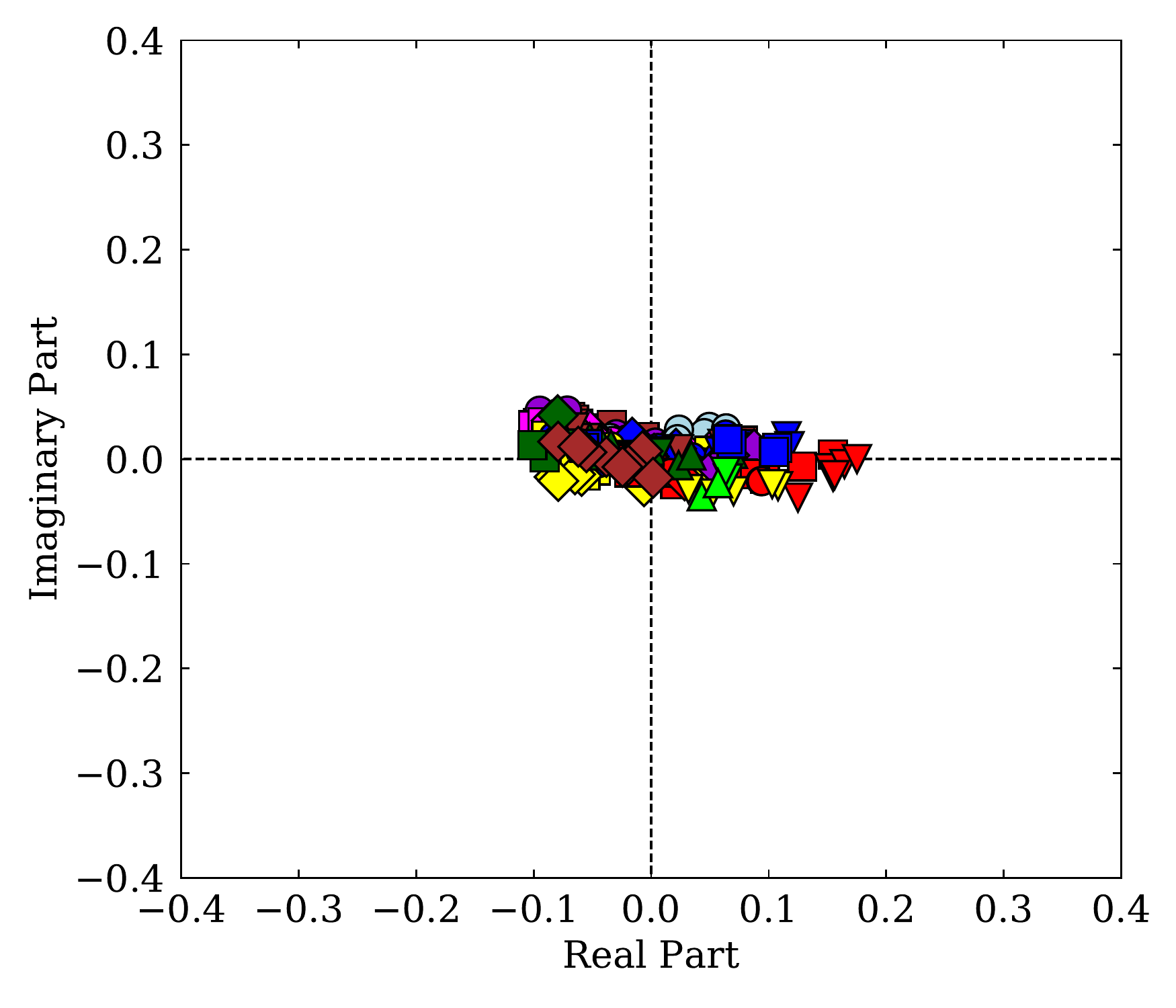}}
	\subfigure[ Eigenvector corresponding to the third largest eigenvalue.]{\includegraphics[width=0.49\textwidth]{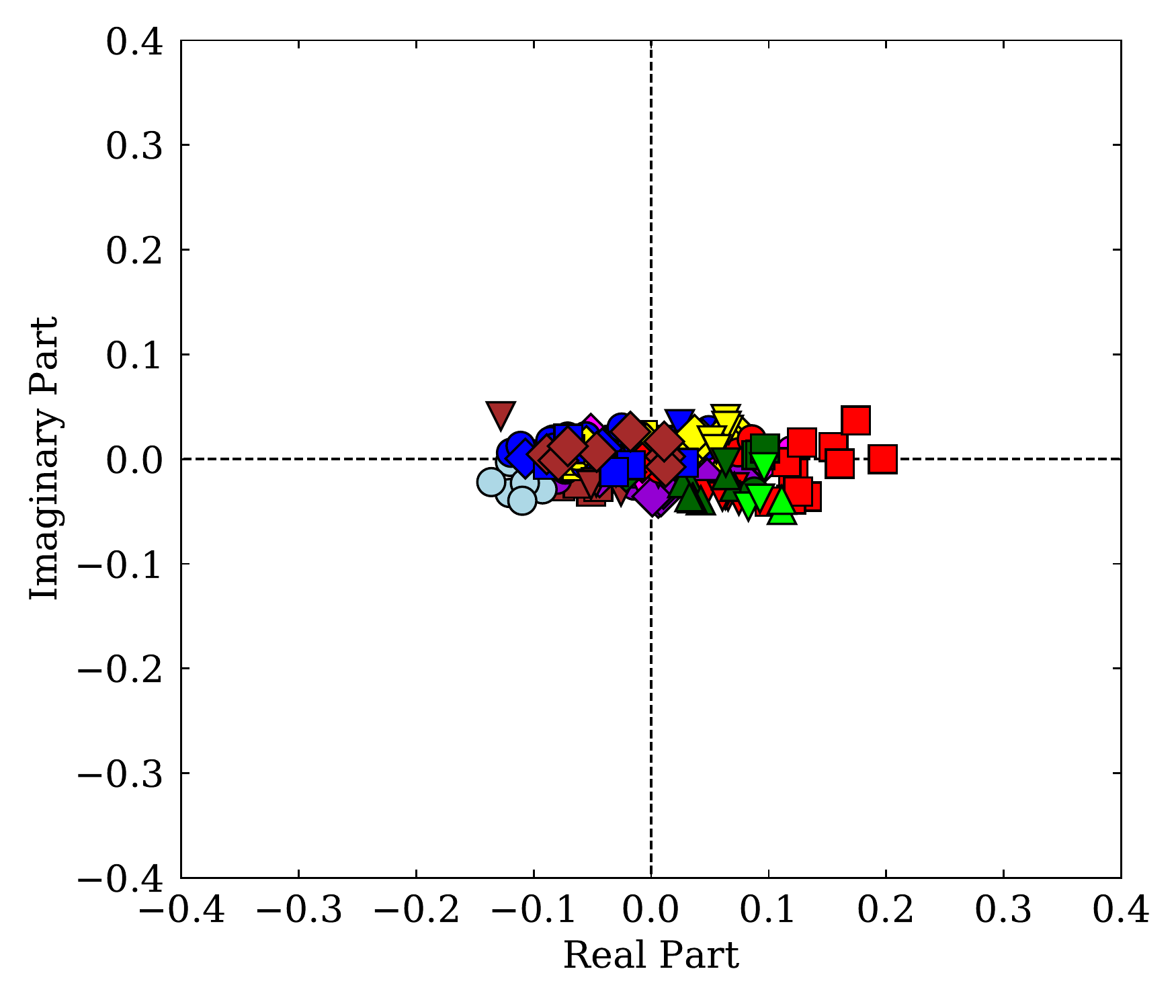}}
	\subfigure[ Eigenvector corresponding to the fourth largest eigenvalue.]{\includegraphics[width=0.49\textwidth]{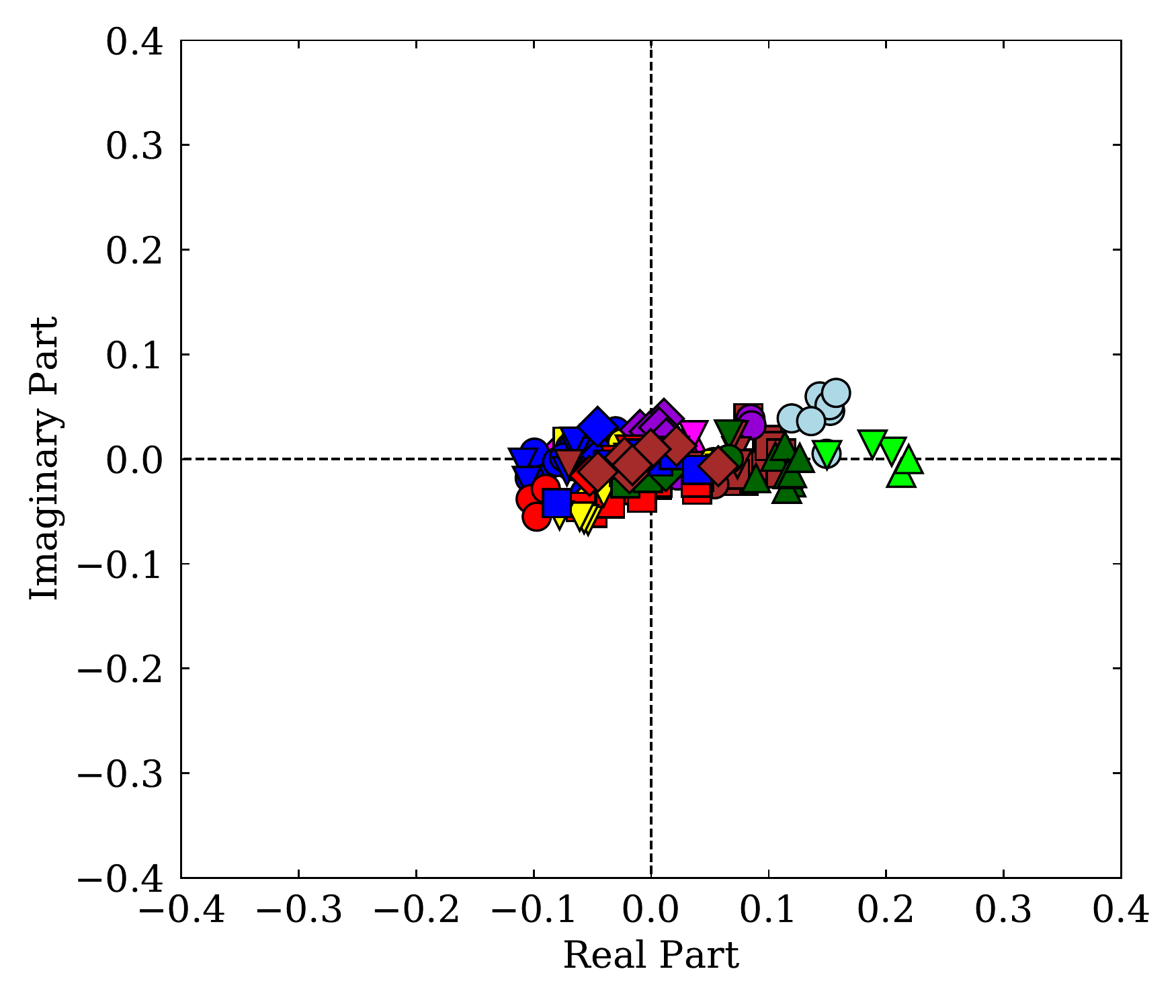}}
	\subfigure[ Eigenvector corresponding to the fifth largest eigenvalue.]{\includegraphics[width=0.49\textwidth]{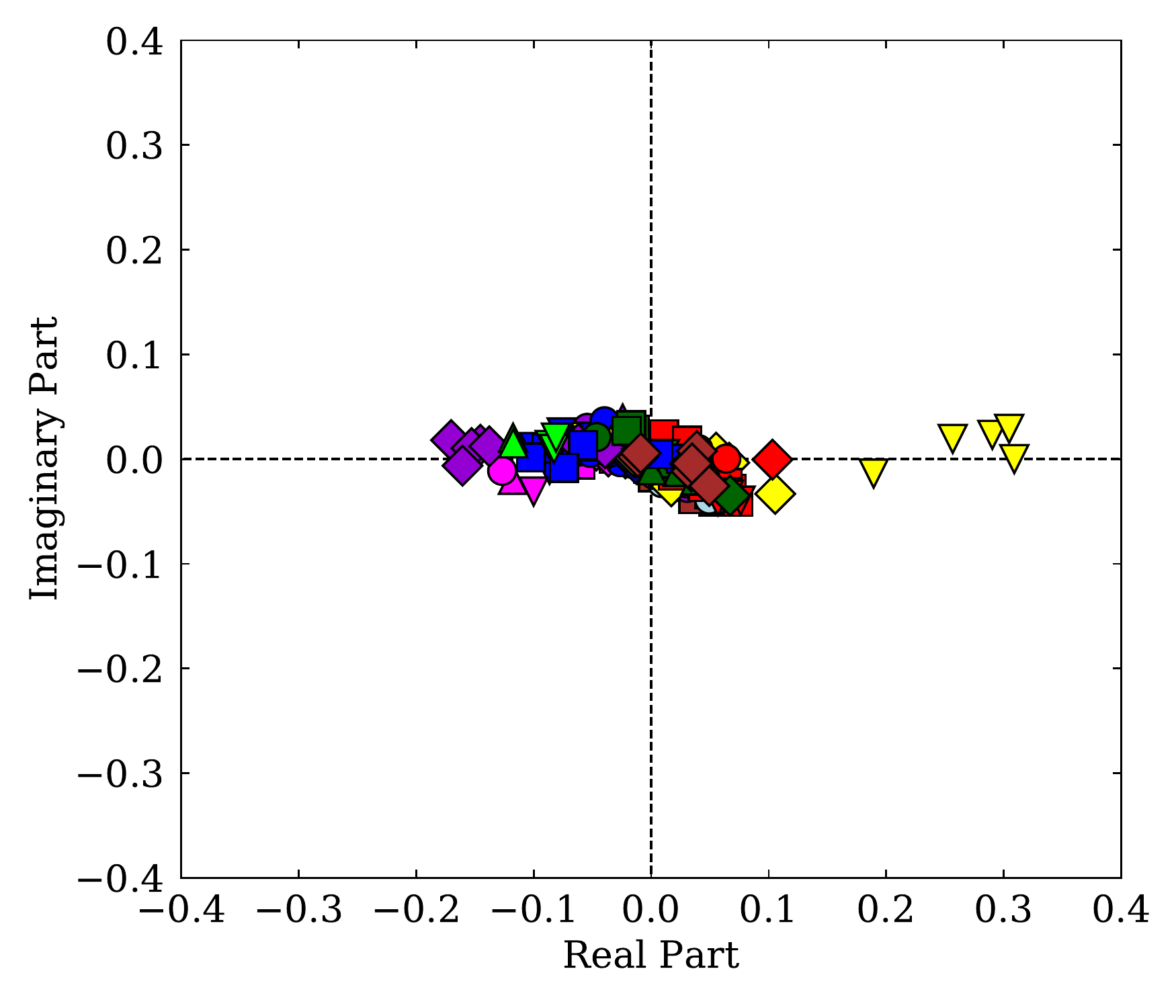}}
	\subfigure[ Eigenvector corresponding to the sixth largest eigenvalue.]{\includegraphics[width=0.49\textwidth]{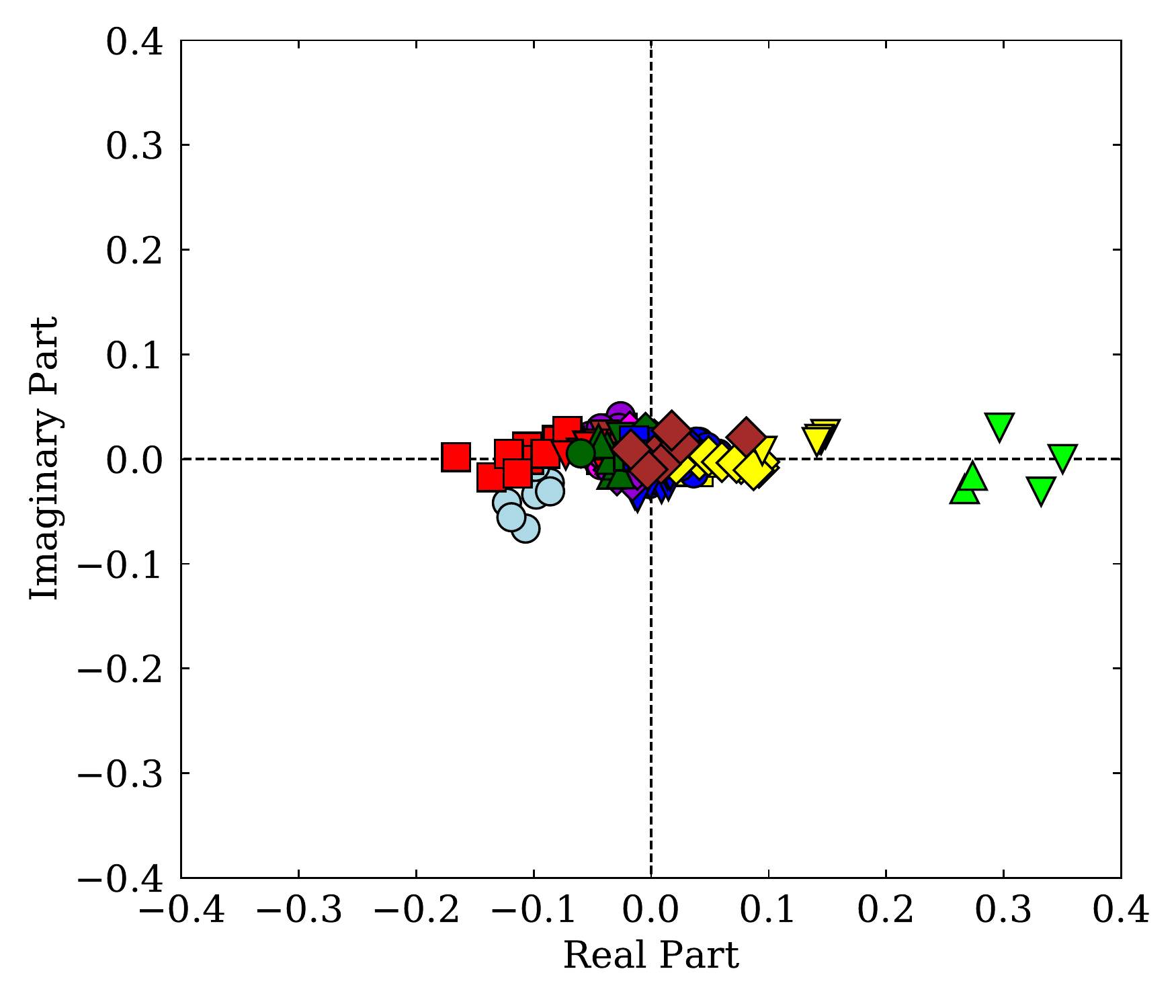}}
    \caption{Immediate eigenvectors' coefficients obtained from complex correlation matrix.}
    \label{fig:eigs_i}
\end{figure}

We start our analysis by looking at the eigenvalues of complex correlation matrix obtained for Nikkei 225 stock quotations in 2014.
As it is shown in Fig. \ref{fig:eigen_hist}, there is one extremely large eigenvalue, few relatively big and the rest form a bulk.
This structure is in accordance with previous results on financial data correlations \cite{laloux1999noise,plerou1999universal,utsugi2004random} and our approach is able to reproduce them.
It was often stated that the largest eigenvalue describes the so called \textit{market mode}, which represents the overall movement of the market.
This movement was often consistent with the most important index in the analysed market.
Moreover, it was shown in \cite{friedman1981interpreting} that this eigenvalue, when all the correlations are positive, is simply a linear function of the average correlation.
In our case correlations are complex but we checked that normal correlations are positive also for the data used in this work.
To find out, whether the same interpretation holds for complex correlations, let us look at the first eigenvector, corresponding to the largest eigenvalue.
Fig. \ref{fig:eigs_i} shows in panel a) that all the coefficients of the eigenvector, represented with complex points, group together on a small, nearly real and positive section in the complex plane.
Each point corresponds to a certain stock and their colors and shapes represent sectors as shown in Table \ref{tab:colors}.
There is no sectoral structure, maybe apart from few material stocks being closer to zero value.
If we recall that the coefficients of each eigenvector are actually correlations between a particular stock and a principal component, we see that all the stocks are positively correlated with this component and there are almost no delays, since phase difference is close to zero for all the points.
This is exactly what we would expect from the market mode, which should lead all the stocks in a similar, positive and immediate manner.

The few big eigenvalues, apart from the largest one, are commonly associated with market sectors and said to contain significant information about the market structure.
Finally, the rest of eigenvalues, being part of the bulk, are supposed to represent market noise.
Our analysis of the complex correlation matrix shows that the spectral structure of financial high frequency data is probably much richer and complex.
We observe that there are at least three significantly different groups of eigenvectors.
First group consists of few largest eigenvalues, the ones that according to random matrix theory should contain most of the non-noise information.
We called this group the \textit{immediate} components and they are all shown in Fig. \ref{fig:eigs_i}.
The word immediate comes from the fact that all of them have eigenvectors' coefficients spread across the real axis with their phases being approximately equal to zero.
This means that these components influence and are influenced by all the stocks with neglectable delay.
Apart from the first eigenvector, we can also see that stocks from the same sectors tend to group close to each other, which is in line with the traditional interpretation of the components being related to market sectors.

\begin{figure}
    \centering
	\subfigure[ Eigenvector corresponding to the seventh largest eigenvalue.]{\includegraphics[width=0.49\textwidth]{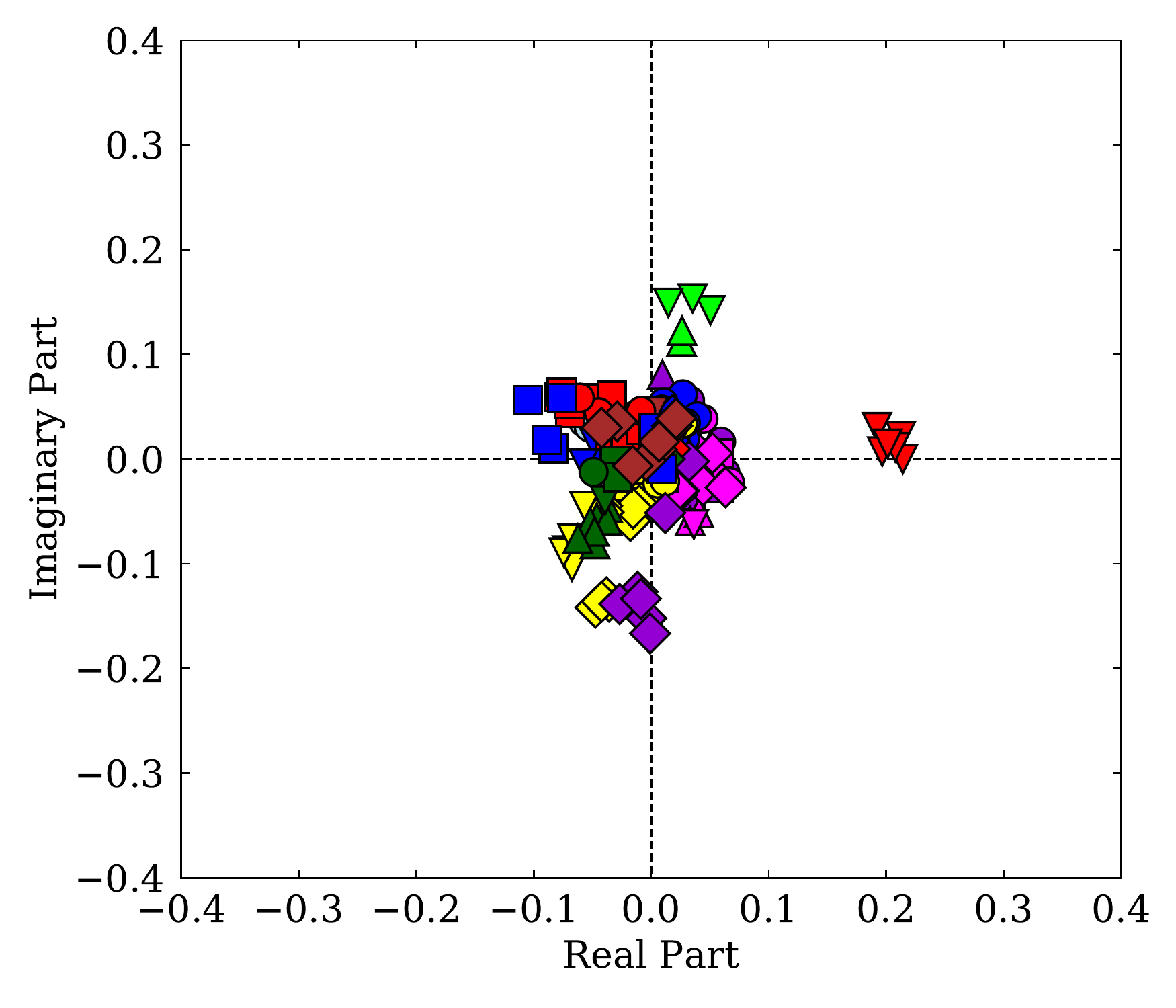}}
	\subfigure[ Eigenvector corresponding to the tenth largest eigenvalue.]{\includegraphics[width=0.49\textwidth]{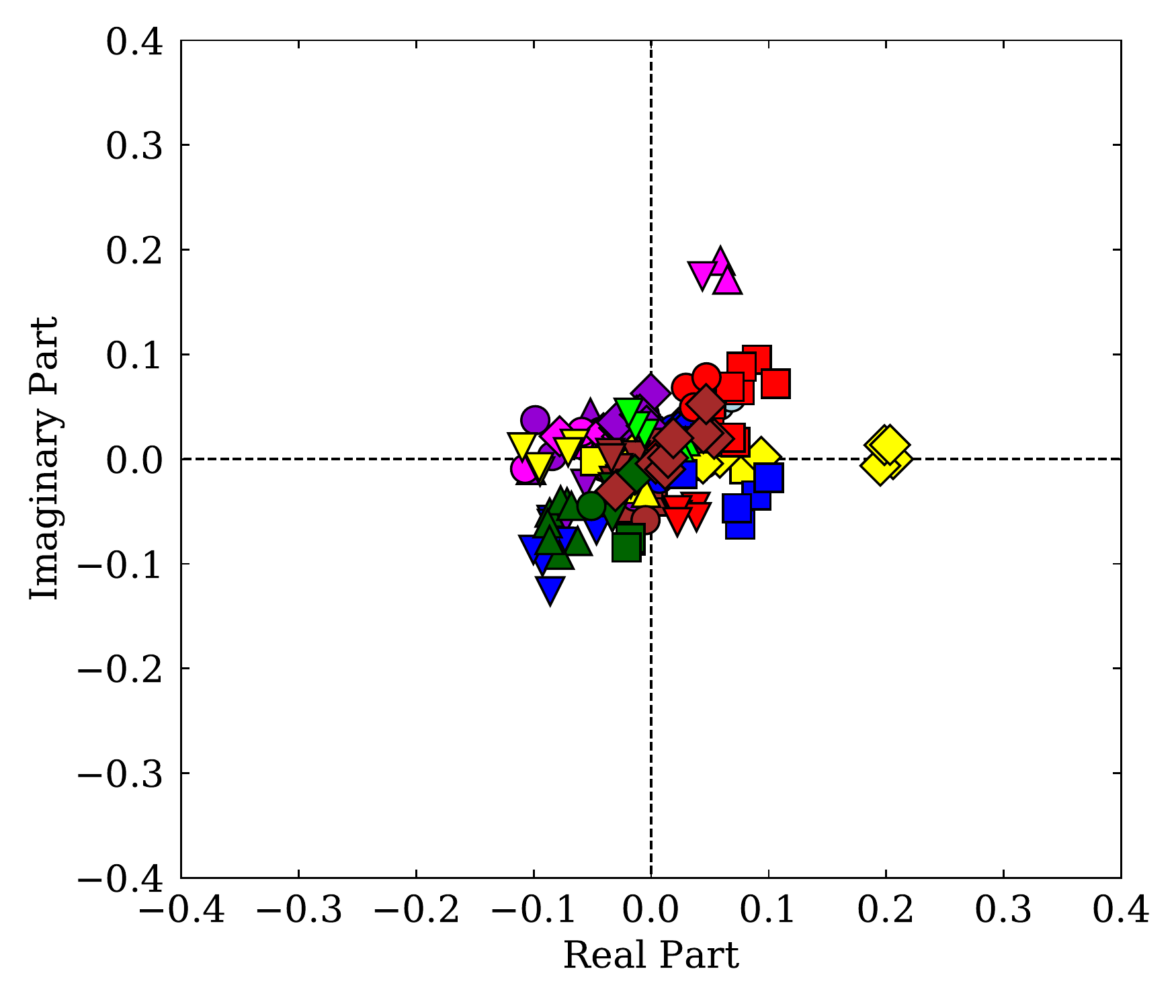}}
	\subfigure[ Eigenvector corresponding to the thirteenth largest eigenvalue.]{\includegraphics[width=0.49\textwidth]{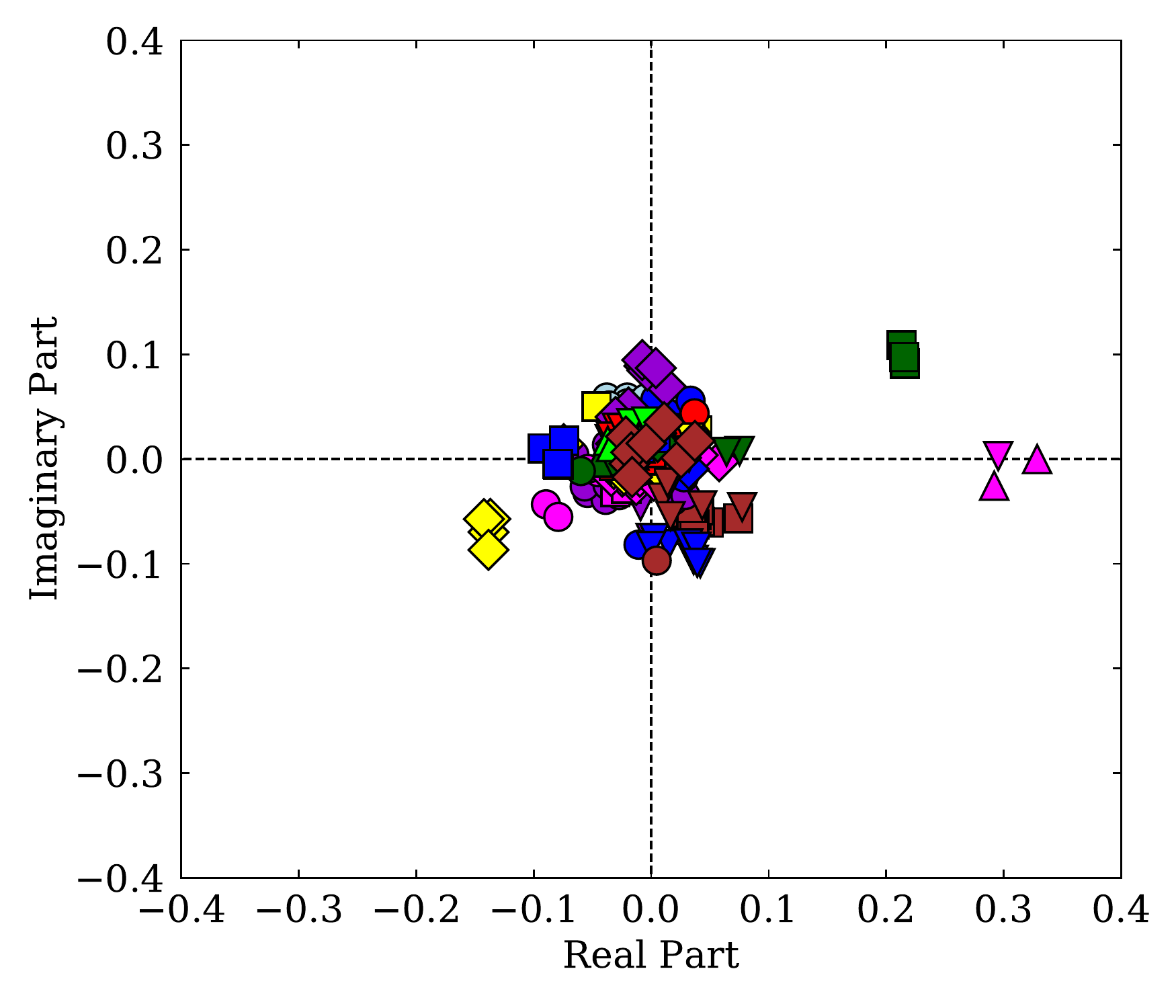}}
	\subfigure[ Eigenvector corresponding to the sixteenth largest eigenvalue.]{\includegraphics[width=0.49\textwidth]{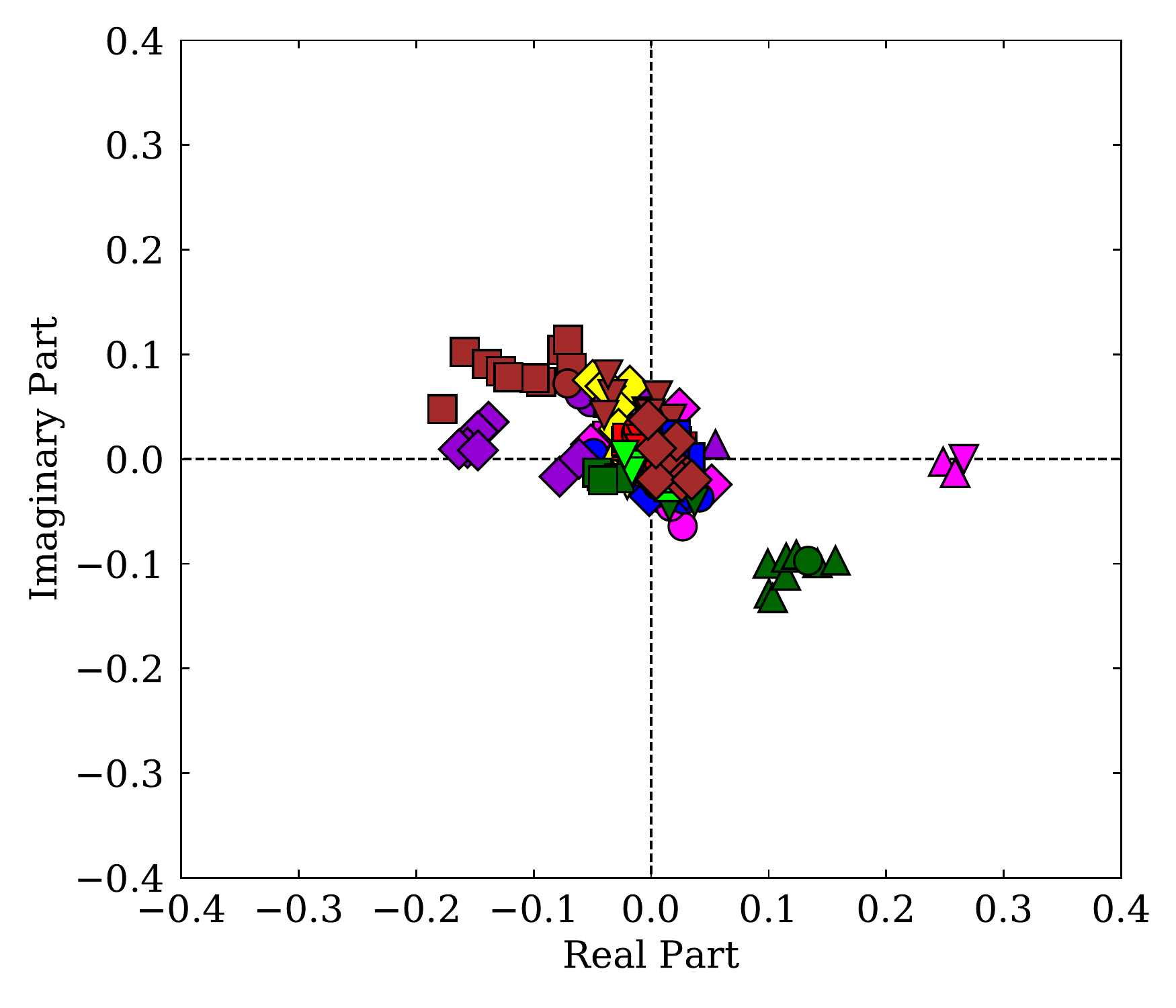}}
	\subfigure[ Eigenvector corresponding to the twentieth largest eigenvalue.]{\includegraphics[width=0.49\textwidth]{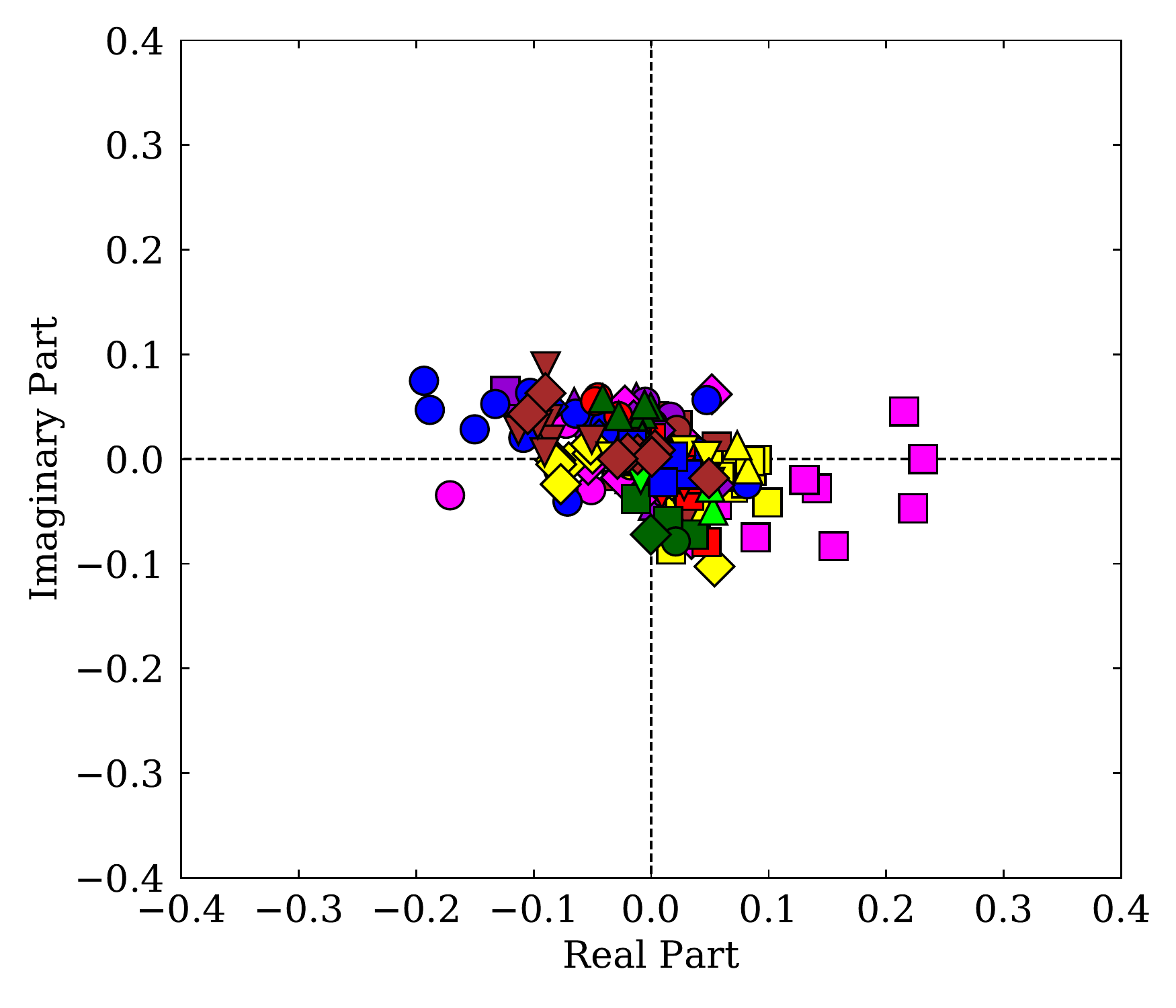}}
	\subfigure[ Eigenvector corresponding to the twenty-fourth largest eigenvalue.]{\includegraphics[width=0.49\textwidth]{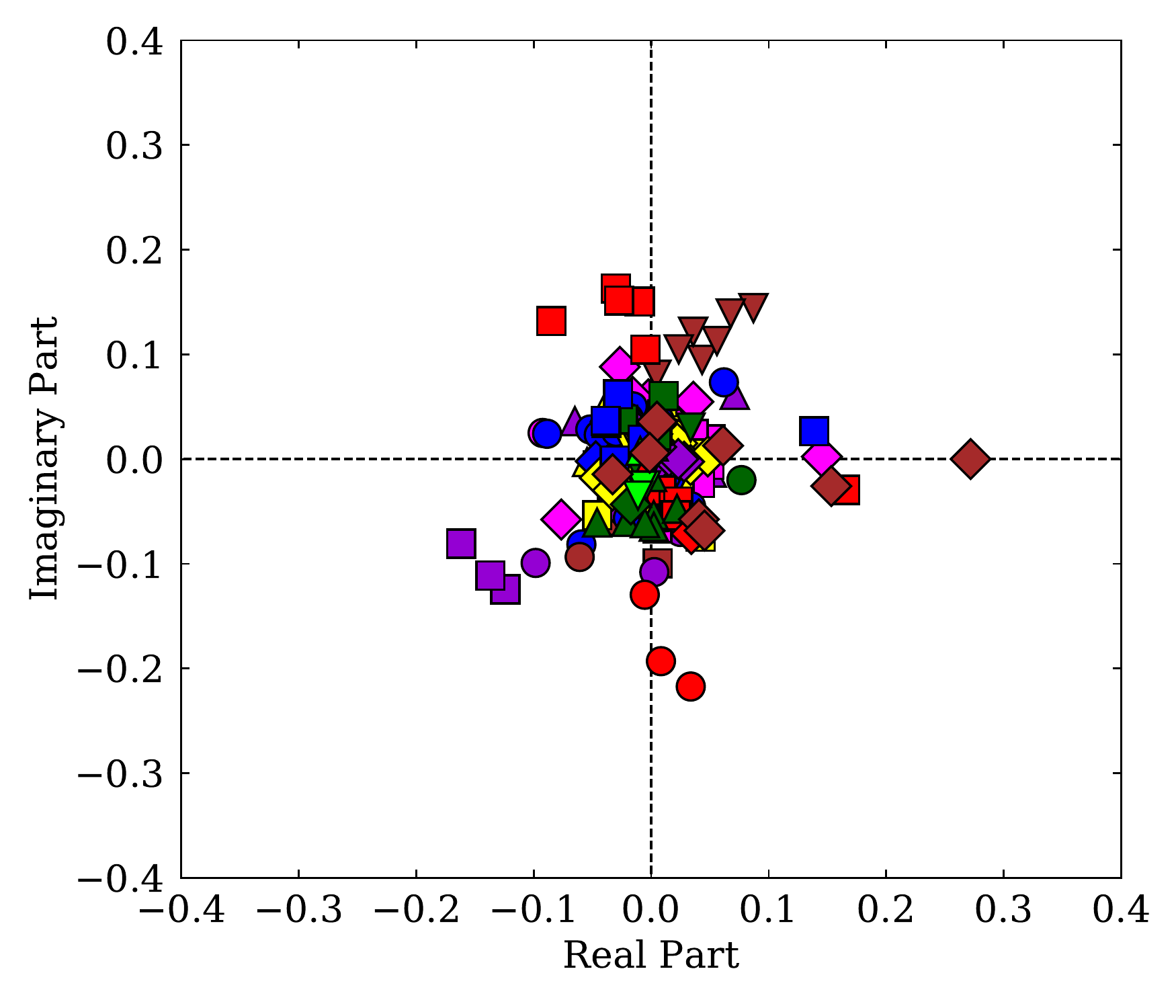}}
    \caption{Delayed eigenvectors' coefficients obtained from complex correlation matrix.}
    \label{fig:eigs_d}
\end{figure}

If we look at further eigenvalues and corresponding eigenvectors, the situation gets even more interesting.
From seventh to somewhere around twenty-fifth eigenvector we can still observe stocks grouping into sectors.
This time, however, they are no longer spread only on the real axis.
Some of these components, which we called \textit{delayed}, are presented in Fig. \ref{fig:eigs_d}.
There is always few stocks from one sector or subsector occupying higher positive correlations on the real axis.
We see them as being the core or the drivers of a given component.
For example the cores of seventh and tenth components are respectively Financials and Capital Goods (specifically insurance and construction).
Then we can see one or more groups of subsectors that are leading or delayed, i.e. have positive or negative phase.
Is there an explanation for these lead lag relations?
We believe that they are not accidental, just like stocks being close to each other as a result of belonging to the same sector.
To prove that, lets take a closer look at the delayed eigenvectors, starting with the seventh component shown in panel a) of Fig. \ref{fig:eigs_d}.
The main driver of this component is the financial sector, with its stocks having the largest correlation magnitude and having a nearly zero phase difference.
After a closer look, we see that this component separates not only sectors but even subsectors.
All the financial stocks on a far positive real axis are from the insurance subsector.
Moreover, these are all the insurance companies in the Nikkei 225 index.
The other financial institutions are not correlated or even negatively correlated with this component.
Other groups highly correlated with this component are the gas and electricity subsectors.
They seem to be leading this component, which might be surprising since it is represented mainly by insurance.
We suspect, however, that this component is connected to households and their prosperity.
That is why it is led by gas and electricity, and is positively correlated with the insurance subsector.
Next three components shown in panels b), c) and d) are all strongly connected to all the stocks from petroleum and mining sectors, specifically oil and coal products.
These subsectors are of a great importance in Japan which lacks fossil fuels and needs to import them from other countries mainly through sea transportation.
We argue that these complex principal components represent fossil fuels and their dependencies across the market.
The tenth component is highly correlated with construction and depicts its lagged relation to mining and petroleum.
Both thirteenth and sixteenth eigenvectors are driven by the fossil fuels.
The first one follows the marine transport which represents the main source of fossil fuels.
The second one, on the other hand, is leading the train and bus transport, which is dependent on petrol.
The eigenvectors shown in panels e) and f) are more chaotic but we can still see stocks of the same sector being close to each other.
These delayed eigenvectors, clearly linked to financial and economic dependencies, are in the bulk which meant to be mostly noise.
The above findings confirm the predictions of \cite{lillo2005spectral}, which showed that in case of factor models, bulk does not need to be driven only by the statistical uncertainty.

\begin{figure}
    \centering
	\subfigure[ Eigenvector corresponding to the forty-first largest eigenvalue.]{\includegraphics[width=0.49\textwidth]{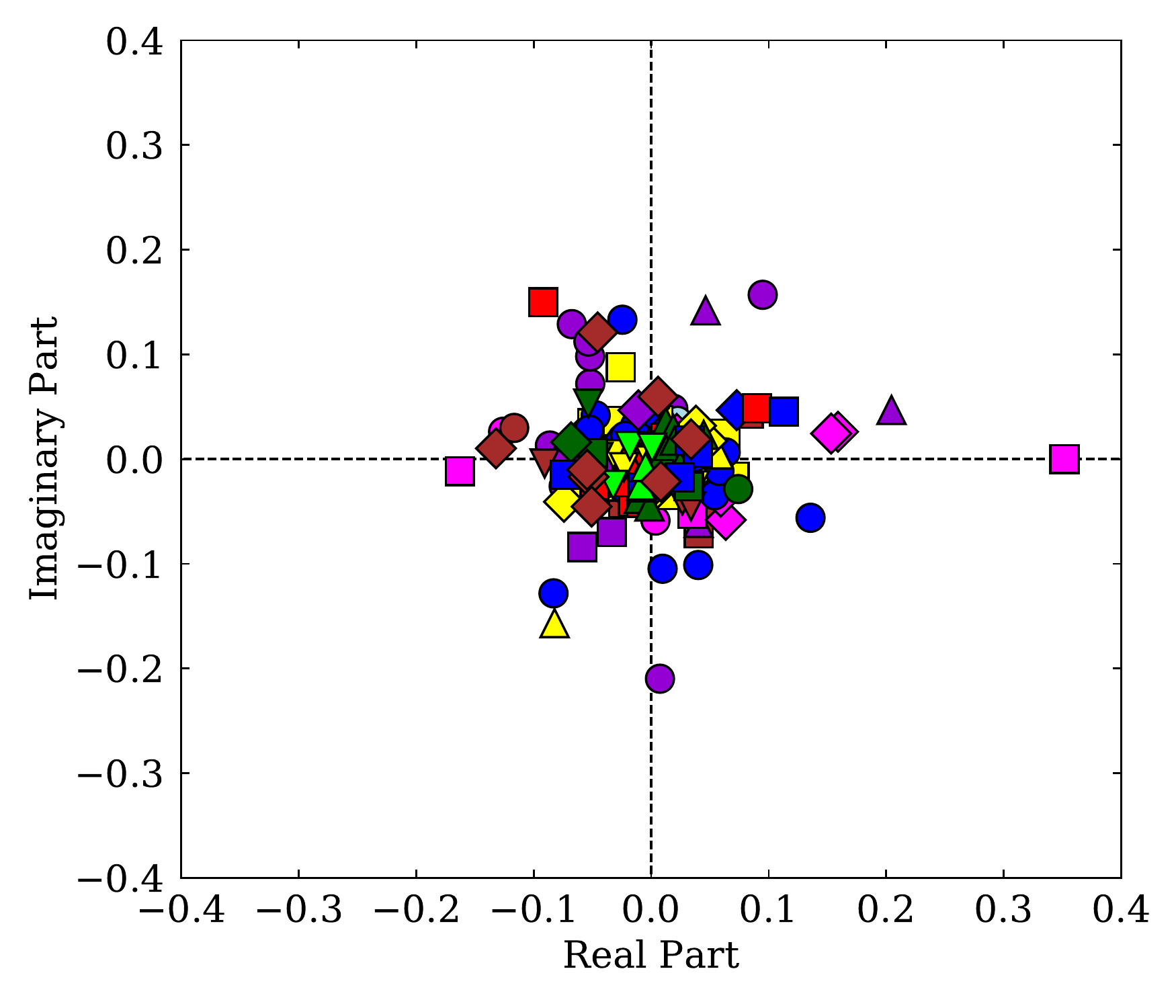}}
	\subfigure[ Eigenvector corresponding to the fifty-sixth largest eigenvalue.]{\includegraphics[width=0.49\textwidth]{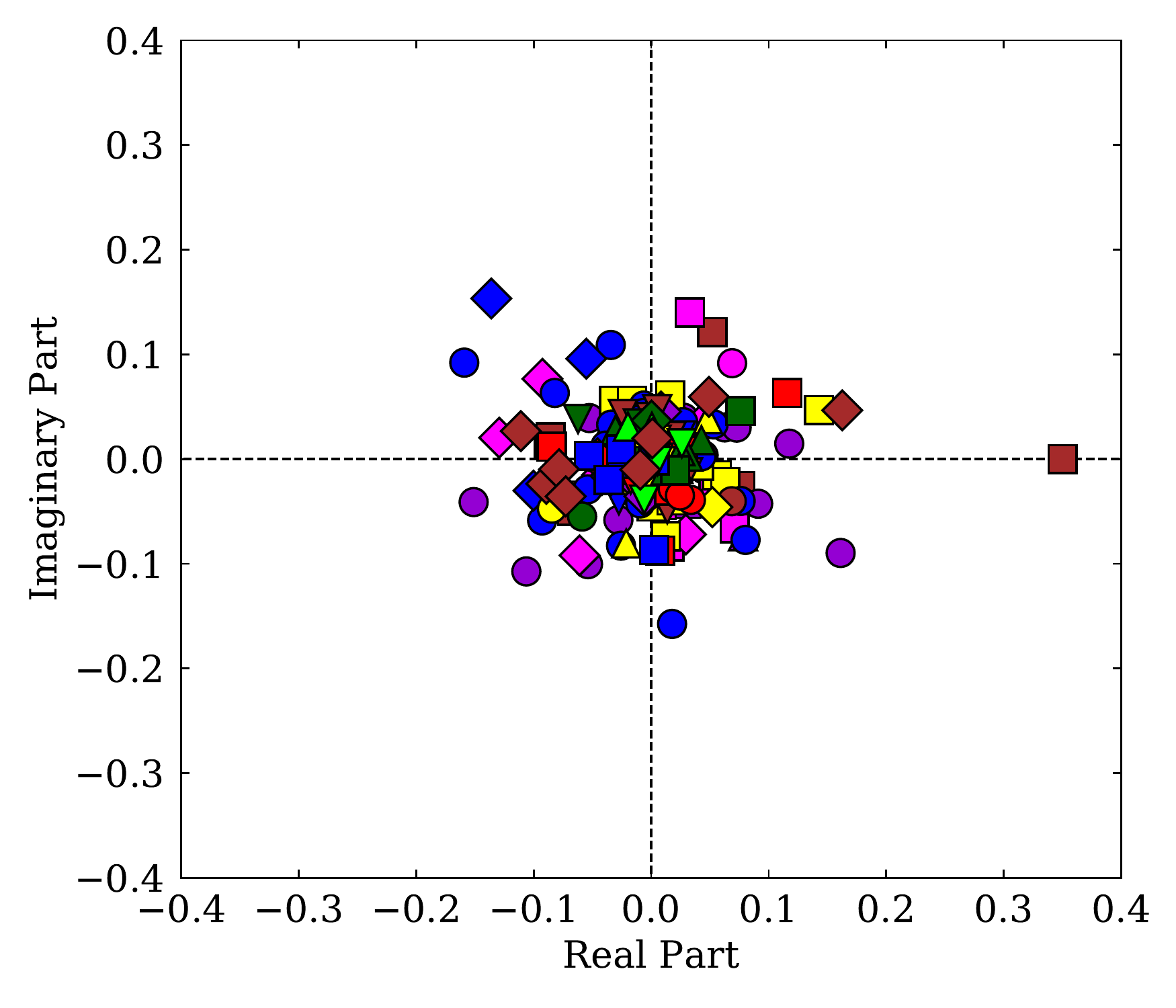}}
	\subfigure[ Eigenvector corresponding to the ninetieth largest eigenvalue.]{\includegraphics[width=0.49\textwidth]{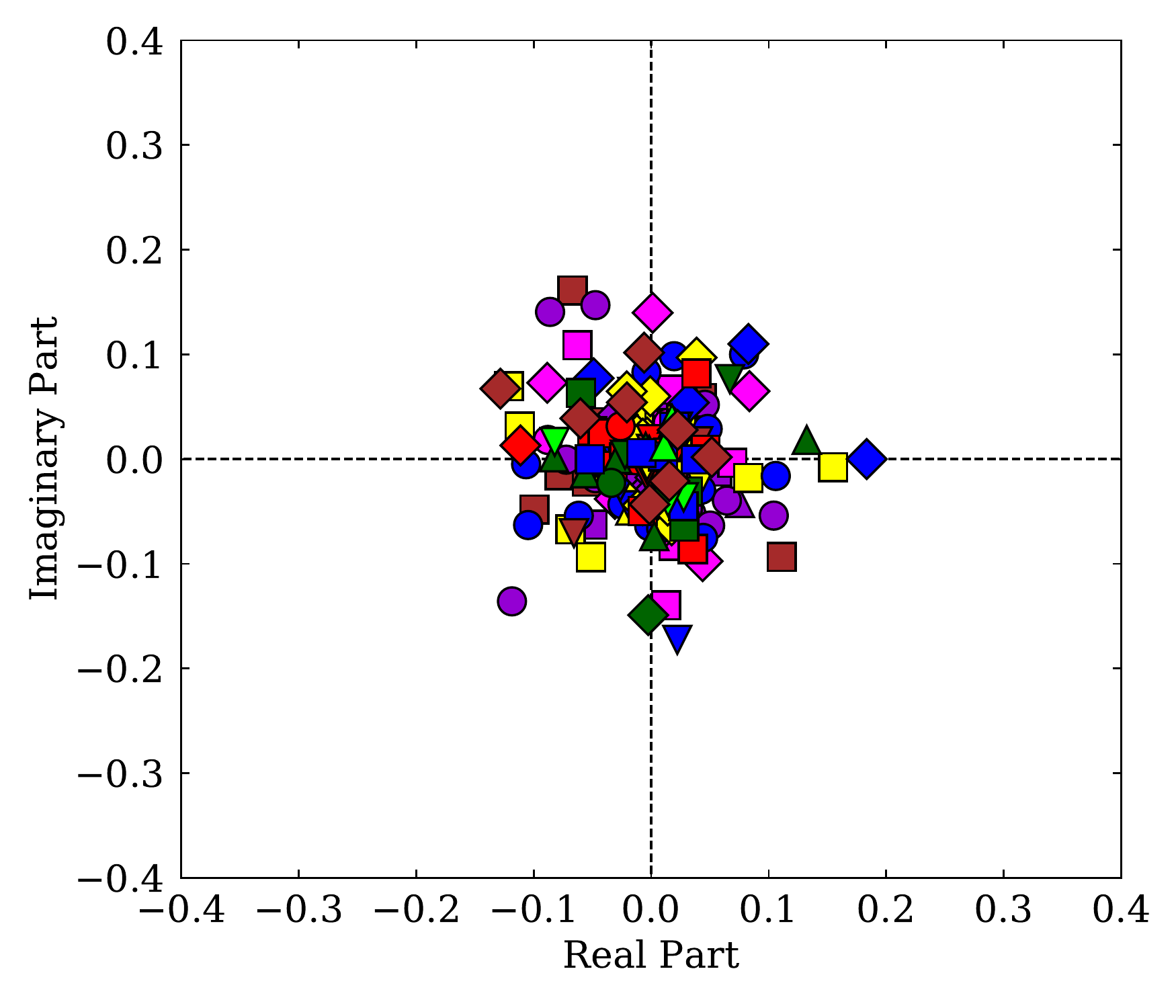}}
	\subfigure[ Eigenvector corresponding to the last eigenvalue.]{\includegraphics[width=0.49\textwidth]{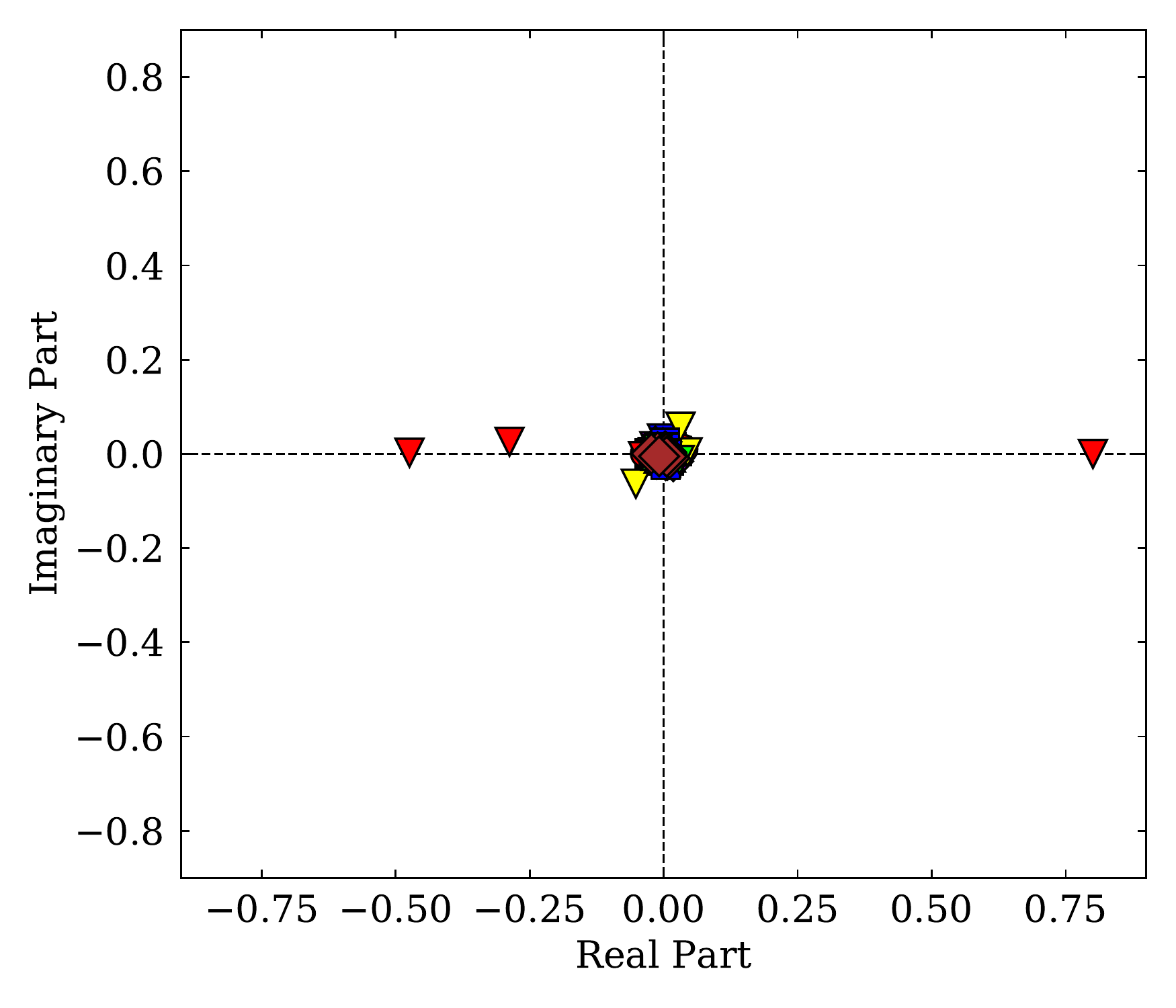}}
    \caption{Chaotic eigenvectors' coefficients obtained from complex correlation matrix.}
    \label{fig:eigs_c}
\end{figure}

The last group consists of \textit{chaotic} components.
As shown in Fig. \ref{fig:eigs_c}, it is difficult to find any, at least connected to sectors, structures among their complex coefficient.
Nevertheless, we would like to comment on two interesting phenomena that we observed.
First, for the vast majority of these components we observe that there is one stock, that may be different for different eigenvectors, which has the highest correlation magnitude and at the same time its phase is equal to zero.
This behavior is shown in both panel a) and b) of Fig. \ref{fig:eigs_c} but as said before, it is very common among chaotic components.
An example of an eigenvector without this kind of driving stock is shown in panel c) of Fig. \ref{fig:eigs_c}.
We attribute this effect to mathematical constraints imposed by PCA, namely the orthogonality of components and their decreasing variance.
Second observation is connected to the last eigenvector, shown in panel d) of Fig. \ref{fig:eigs_c}, where there are three stocks with significantly high absolute correlation.
These stocks are all from the same sector and they are all near the real axis, whereas all other stocks group closely to the $(0,0)$ point.
Closer look at these three stocks shows that they not only belong to the same sector but also have a closely related price formation\footnote{Specifically, they are: 8630 - SOMPO Holdings Inc, 8725 - Ms\&Ad Insurance Group Holding Inc and 8766 - Tokio Marine Holdings Inc.}.
Therefore, this component is a linear combination of very similar time series, with combination coefficients which sum up to somewhere around zero.
As a result it has a very low variance and again this is connected to mathematical constraints of PCA.

\begin{figure}
    \centering
	\subfigure[ Maximum Spanning Tree.]{\includegraphics[width=0.90\textwidth]{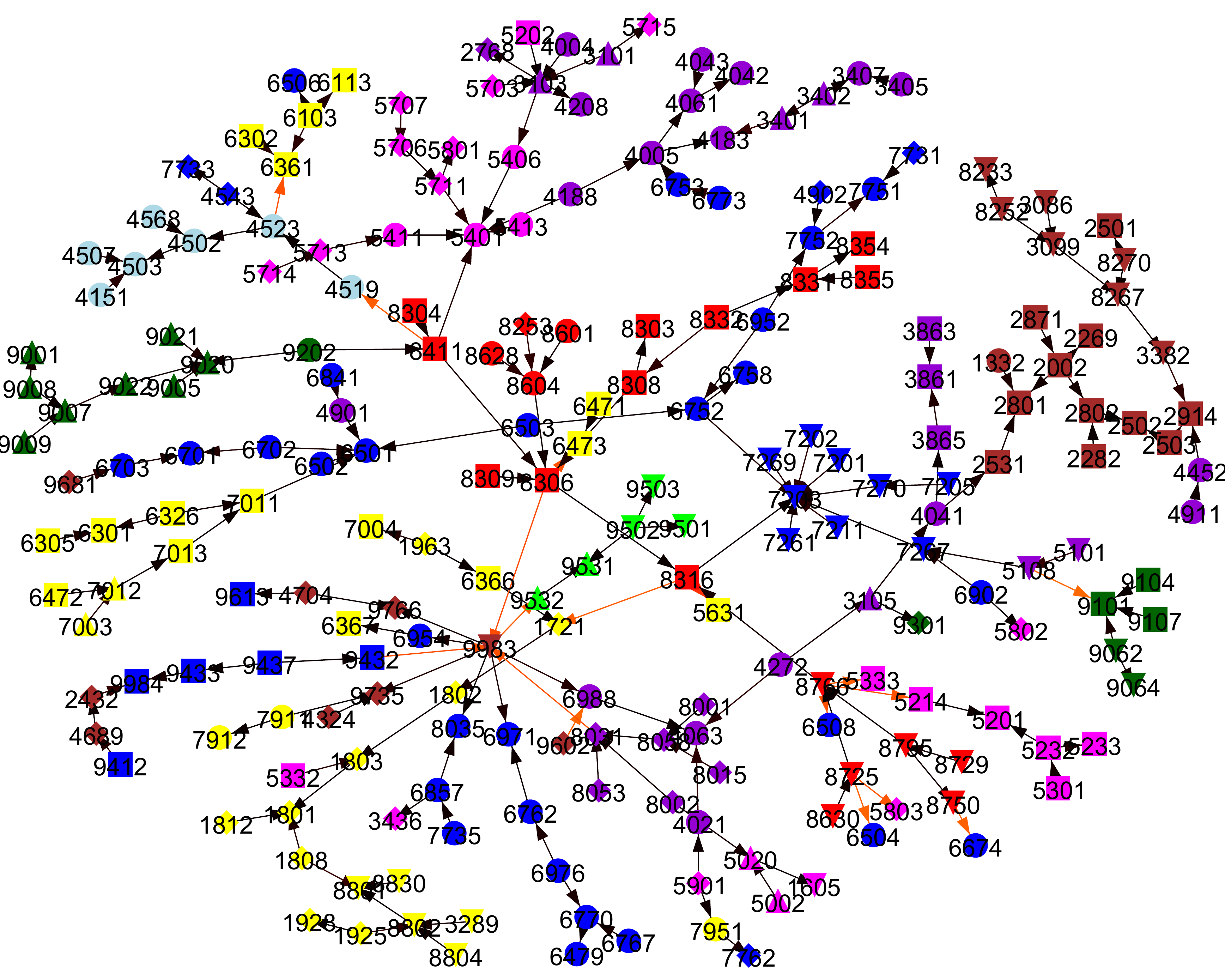}}
	\subfigure[ Planar Maximally Filtered Graph.]{\includegraphics[width=0.90\textwidth]{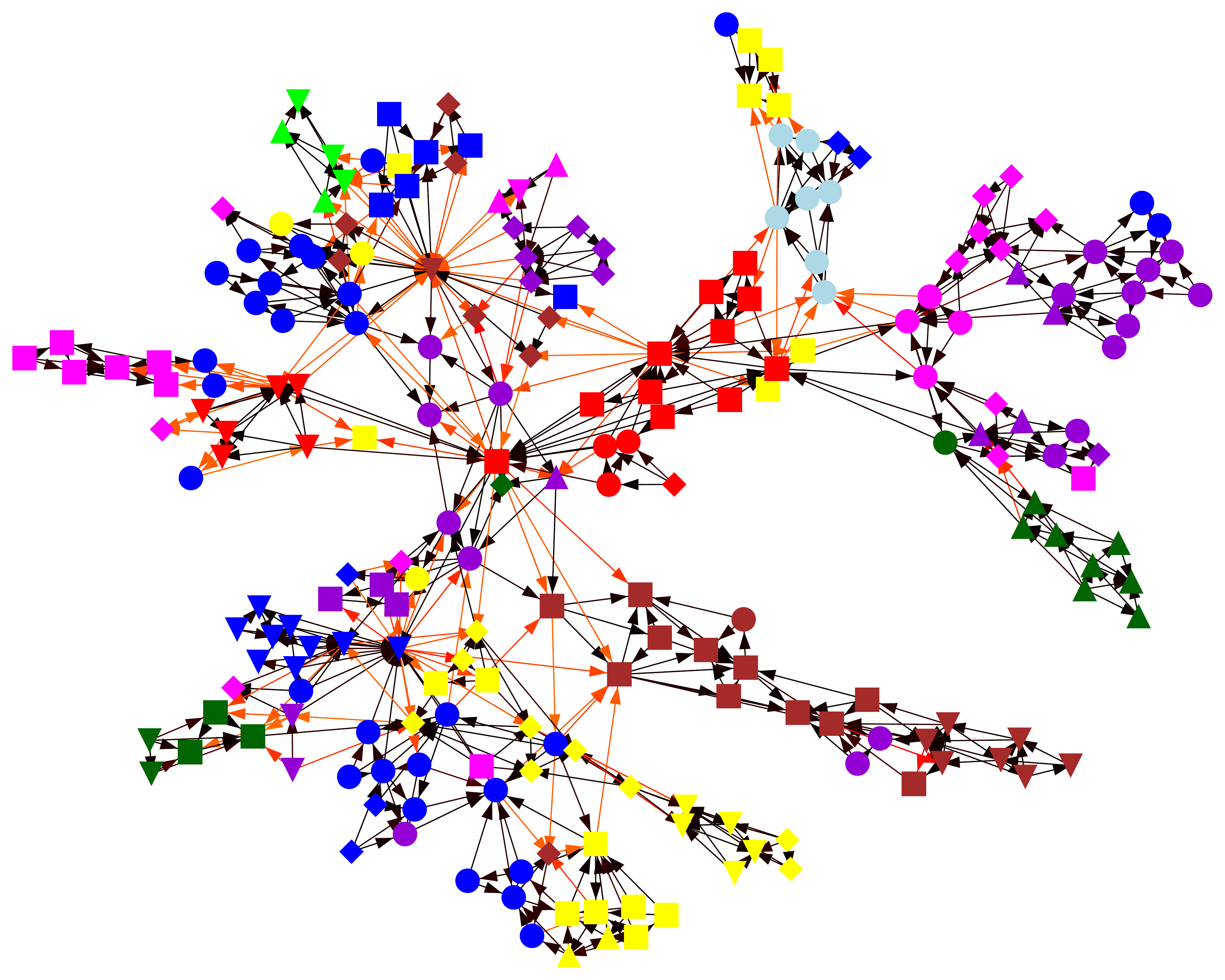}}
    \caption{Filtered complex correlation graphs for TSE intraday data (2014). Nodes represent stocks from Nikkei 225 index and their colors represent sectors. Arrows indicate which stock is leading the other and their color corresponds to the phase difference between these stocks.}
    \label{fig:mst_pmfg}
\end{figure}

As a next step, we would like to show aggregated results, instead of analysing each eigen subspace separately.
We suspect the complex correlation matrix to code information about sectors, in accordance with what we have seen so far.
At first, we shall get rid of the market mode by using the sum from eq. (\ref{eq:decom}) but without the first element ($i = 1$).
Then, we use a classical filtering method called the \textit{Minimum Spanning Tree}, which was used considerable number of times with financial data \cite{mantegna1999hierarchical,iori2007weighted,onnela2003dynamics,bonanno2001high,wilinski2013structural}.
Because our correlations are complex, we cannot use the usual formula for the distances in the correlation network.
Instead, we will take correlations magnitude $s_{kl} = | \rho_{kl} |$ as the edge weight in the filtering algorithm, and calculate the \textit{Maximum Spanning Tree} (MST), since we want to maximize correlations.
Additionally, we will use the phase $\theta_{kl}$ in order to determine the direction of an edge.
If $\theta_{kl} > 0$ ($\theta_{kl} < 0$) then the edge goes from $l$ to $k$ ($k$ to $l$) and latter leads the previous one.
Result of this procedure is shown in Fig. \ref{fig:mst_pmfg} panel a), with four digits numbers being the stocks tickers at TSE.
We observe a significant sector and subsector clustering, which again confirms that our method is consistent with previous findings obtained by traditional correlation methods \cite{mantegna1999hierarchical,iori2007weighted,bonanno2001high}.
Moreover, there is an additional outcome, coded by the edge direction and the phase difference.
Different arrow colors are connected to the phase difference size.
Black corresponds to small phase differences, red means that the phase difference is close to $\pi/2$ and orange is for phase differences around $\pi$ or $-\pi$, which suggests negative correlations.
Connections between stocks from the same sector are all black, whereas many intersector edges are orange.
This indicates that inside of a sector, stocks tend to follow each other rather closely.
On the other hand, there are strong, but probably negative correlations between stocks from different sectors.
Another interesting observation is that highly connected stocks have much more arrows pointing at them than pointing outside.
That means that they are actually leading the rest of the sector.
To point out few examples, we see this for stocks: 7203 (Automobiles and Auto parts), 5401 (Steel products) and 8031 (Trading companies).
This result is consistent with a frequent remark that the highly connected stocks in filtered correlation networks are actually the most significant ones.
Similar results and conclusions were also obtained by estimating the time-dependent correlation function \cite{huth2014high,kullmann2002time}.
A different outcome of lead lag analysis may be found in \cite{curme2015emergence}, but in this case all the correlation were analysed, without any prior data filtering.

In order to valid the statements made above, we also used a less restrictive filtering method, that leaves three times more connections but has the MST backbone among them.
This method is called Planar Maximally Filtered Graph (PMFG) \cite{tumminello2005tool} and was often used as a filtering method for stock market dependencies \cite{buccheri2013evolution,fiedor2014networks}.
PMFG in Fig. \ref{fig:mst_pmfg} panel b), drawn without tickers to make the graph more readable, confirms the observations made on MST.
Furthermore, this filtering method allow cliques and they are often formed by stocks belonging to the same sectors, similarly to what was shown in \cite{tumminello2005tool}.

A single outlier to all the characteristics presented above, is the stock 9983 - Fast Retailing (Retail).
It is connected to different sectors, it has out-degree higher then in-degree, despite having many connections, and it has both edges with high and low phase difference.
The reason for this stock to have so peculiar and exceptional structure of relations is that it is by far the most influential stock of the Nikkei 225 index.
It had the highest index weight, much higher than any other stock analysed here.

\begin{figure}
    \centering
	\subfigure[ Pairs of stocks from different sectors.]{\includegraphics[width=0.49\textwidth]{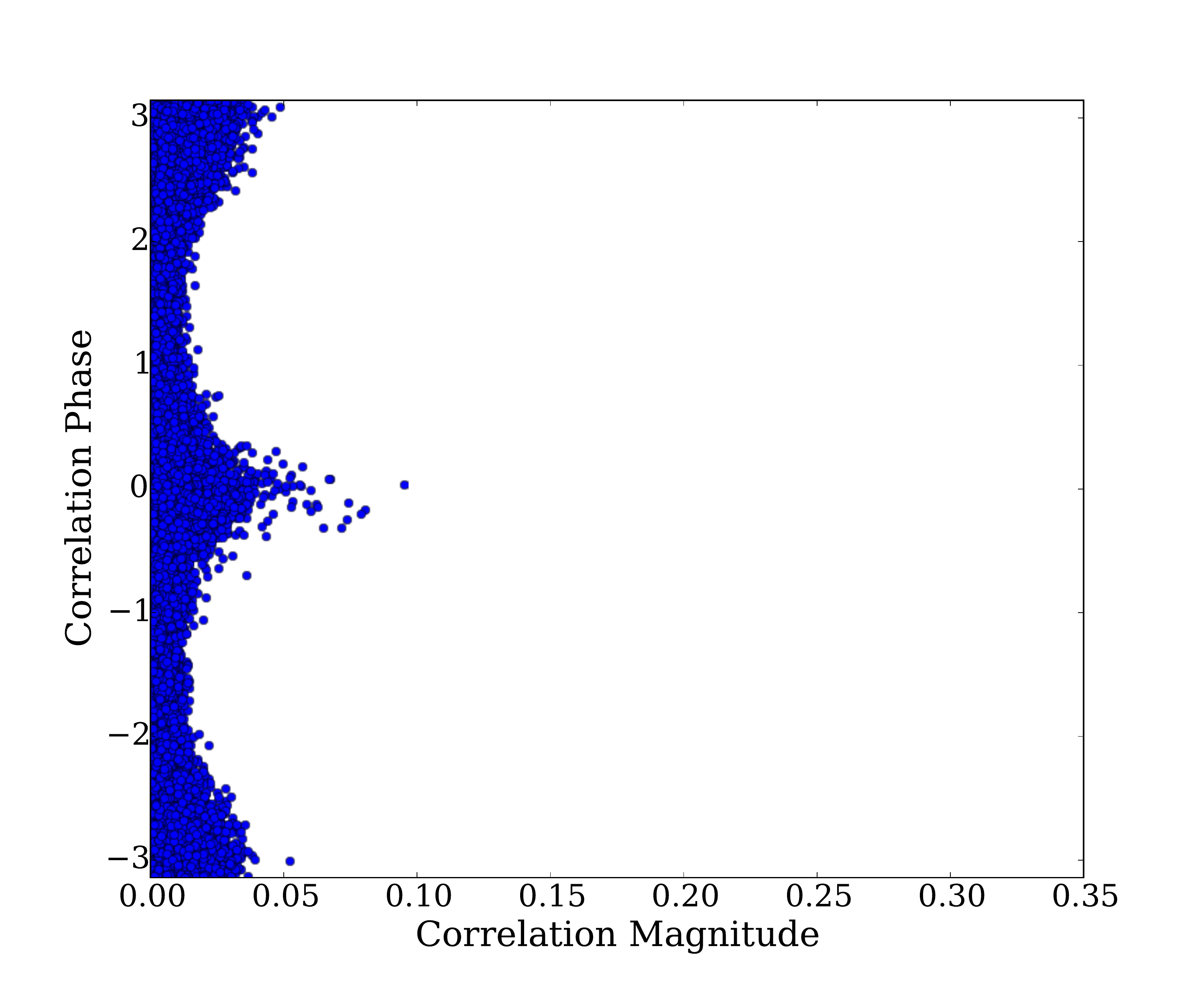}}
	\subfigure[ Pairs of stocks from the same sectors.]{\includegraphics[width=0.49\textwidth]{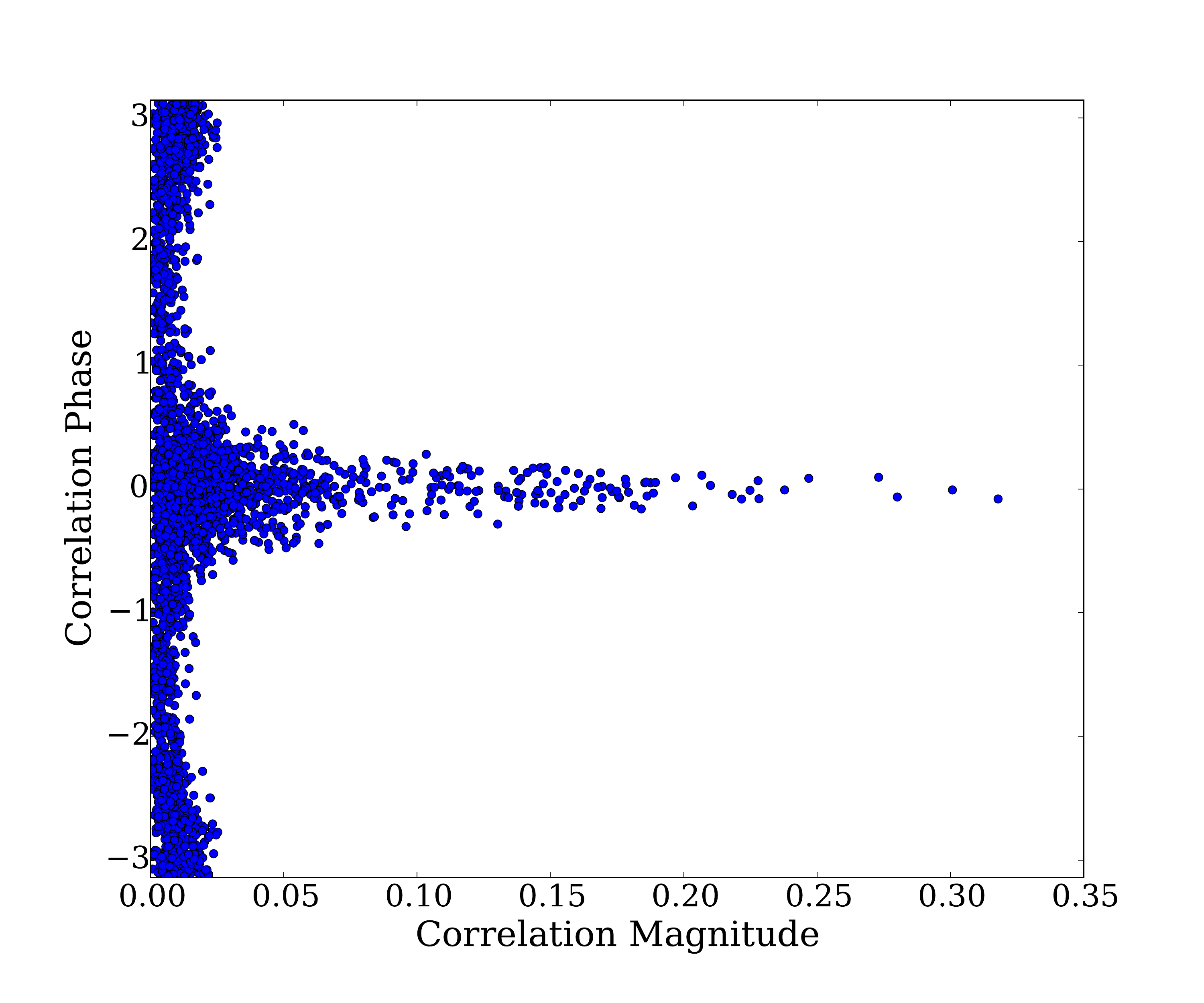}}
    \caption{Dependence between correlation magnitude and its phase for all pairs of stocks in Nikkei 225 index in 2014.}
    \label{fig:corr_phase}
\end{figure}

As a last remark, we present Fig. \ref{fig:corr_phase}, which shows the dependence between the complex correlation magnitude and the complex correlation phase difference.
As we can see in panel b), the highest correlations are between stocks from the same sector, and they are characterized by a small phase differences.
The pairs with large phase differences and more significant correlations, on the other hand, are more likely to be from different sectors.
This is in line with our findings in complex correlation based filtered graphs.

\section{CONCLUSIONS}

We presented a novel approach that combines a Fourier transform based method for calculating correlations in high frequency data, with a Hilbert transform based method of including lead-lag relations in a correlation measure.
This way we propose a unique tool to work with unevenly spaced data.
Moreover, the calculation of the complex correlation matrix from Fourier estimator coefficients is completely analytical and do not require additional numerical approximations.
We should point out, however, that as any other method, it is limited by the data resolution and affected by the finite size effects.

We further used this approach with TSE intraday data and proved that it gives insightful information, especially when analysing eigenspace of the complex correlation matrix.
We confirmed the dominance of sector relations in the market structure and showed that they influence also the bulk eigenvalues.
Furthermore, even though the delays showed by the phase differences are small, they suggest that stocks with higher correlations are leading others.
It is particularly observed for groups of stocks from the same sector and it is in line with the intuition that stocks with higher degree in filtered correlation network are the leading ones.

We believe that this method may and should be used with other types of data, in particular data characterised by unevenly spaced observations.

\begin{table}[t]
\caption{Sectors of Nikkei225 stocks.}
\label{tab:colors}
\begin{tabular}{|c|l|l|}
\hline
\raisebox{-.5\height}{\includegraphics[width=0.5cm]{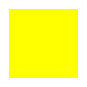}} & Machinery & \multirow{5}{*}{Capital Goods and Others}\\
\cline{1-2}
\raisebox{-.5\height}{\includegraphics[width=0.5cm]{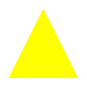}} & Shipbuilding & \\
\cline{1-2}
\raisebox{-.5\height}{\includegraphics[width=0.5cm]{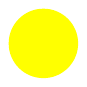}} & Other manufacturing & \\
\cline{1-2}
\raisebox{-.5\height}{\includegraphics[width=0.5cm]{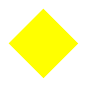}} & Construction & \\
\cline{1-2}
\raisebox{-.5\height}{\includegraphics[width=0.5cm]{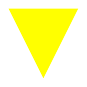}} & Real estate & \\
\hline
\raisebox{-.5\height}{\includegraphics[width=0.5cm]{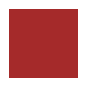}} & Foods & \multirow{4}{*}{Consumer Goods}\\
\cline{1-2}
\raisebox{-.5\height}{\includegraphics[width=0.5cm]{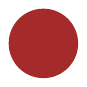}} & Fishery & \\
\cline{1-2}
\raisebox{-.5\height}{\includegraphics[width=0.5cm]{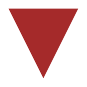}} & Retail & \\
\cline{1-2}
\raisebox{-.5\height}{\includegraphics[width=0.5cm]{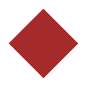}} & Services & \\
\hline
\raisebox{-.5\height}{\includegraphics[width=0.5cm]{Banking.pdf}} & Banking & \multirow{4}{*}{Financials}\\
\cline{1-2}
\raisebox{-.5\height}{\includegraphics[width=0.5cm]{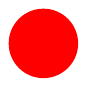}} & Securities & \\
\cline{1-2}
\raisebox{-.5\height}{\includegraphics[width=0.5cm]{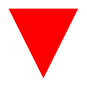}} & Insurance & \\
\cline{1-2}
\raisebox{-.5\height}{\includegraphics[width=0.5cm]{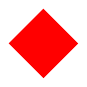}} & Other financial services & \\
\hline
\raisebox{-.5\height}{\includegraphics[width=0.5cm]{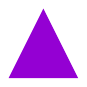}} & Textiles and apparel & \multirow{10}{*}{Materials}\\
\cline{1-2}
\raisebox{-.5\height}{\includegraphics[width=0.5cm]{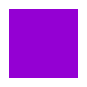}} & Pulp and paper & \\
\cline{1-2}
\raisebox{-.5\height}{\includegraphics[width=0.5cm]{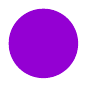}} & Chemicals & \\
\cline{1-2}
\raisebox{-.5\height}{\includegraphics[width=0.5cm]{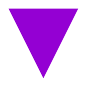}} & Rubber products & \\
\cline{1-2}
\raisebox{-.5\height}{\includegraphics[width=0.5cm]{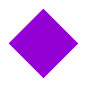}} & Trading companies & \\
\cline{1-2}
\raisebox{-.5\height}{\includegraphics[width=0.5cm]{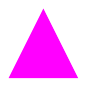}} & Petroleum & \\
\cline{1-2}
\raisebox{-.5\height}{\includegraphics[width=0.5cm]{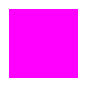}} & Glass and ceramics & \\
\cline{1-2}
\raisebox{-.5\height}{\includegraphics[width=0.5cm]{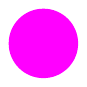}} & Steel products & \\
\cline{1-2}
\raisebox{-.5\height}{\includegraphics[width=0.5cm]{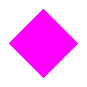}} & Nonferrous metals & \\
\cline{1-2}
\raisebox{-.5\height}{\includegraphics[width=0.5cm]{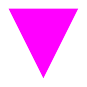}} & Mining & \\
\hline
\raisebox{-.5\height}{\includegraphics[width=0.5cm]{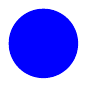}} & Electric machinery & \multirow{5}{*}{Technology}\\
\cline{1-2}
\raisebox{-.5\height}{\includegraphics[width=0.5cm]{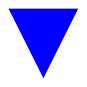}} & Automobiles and Auto parts & \\
\cline{1-2}
\raisebox{-.5\height}{\includegraphics[width=0.5cm]{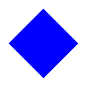}} & Precision instruments & \\
\cline{1-2}
\raisebox{-.5\height}{\includegraphics[width=0.5cm]{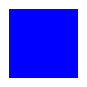}} & Communications & \\
\cline{1-2}
\raisebox{-.5\height}{\includegraphics[width=0.5cm]{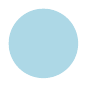}} & Pharmaceuticals & \\
\hline
\raisebox{-.5\height}{\includegraphics[width=0.5cm]{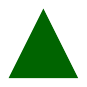}} & Railway and bus & \multirow{7}{*}{Transportation and Utilities}\\
\cline{1-2}
\raisebox{-.5\height}{\includegraphics[width=0.5cm]{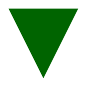}} & Other land transport & \\
\cline{1-2}
\raisebox{-.5\height}{\includegraphics[width=0.5cm]{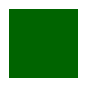}} & Marine transport & \\
\cline{1-2}
\raisebox{-.5\height}{\includegraphics[width=0.5cm]{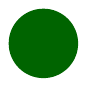}} & Air transport & \\
\cline{1-2}
\raisebox{-.5\height}{\includegraphics[width=0.5cm]{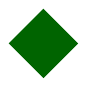}} & Warehousing & \\
\cline{1-2}
\raisebox{-.5\height}{\includegraphics[width=0.5cm]{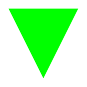}} & Electric power & \\
\cline{1-2}
\raisebox{-.5\height}{\includegraphics[width=0.5cm]{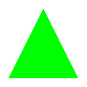}} & Gas & \\
\hline
\end{tabular}
\end{table}

\section{ACKNOWLEDGMENTS}

This work was supported by the National Science Centre under Grant 2015/19/N/ST2/02701.
Authors would also like to thank Tomasz Raducha and two anonymous referees for insightful discussions and suggestions.

\bibliography{bibliography}

\end{document}